\documentclass[11pt]{article}
\usepackage{geometry,amsmath,amssymb}
\numberwithin{equation}{section}
\def\ts{\textstyle}
\def\st{{\textstyle\sum}}

\def\nn{\nonumber}
\def\half{\frac{1}{2}}

\def\p{\partial}
\def\pb{\bar\partial}

\def\a{\alpha}

\def\ad{{\dot\alpha}}
\def\b{\beta}
\def\bd{{\dot\beta}}

\def\t{\theta}
\def\tb{{\bar\theta}}
\def\db{{\bar d}}
\def\uno{1\!\!1}
\def\e{\epsilon}
\def\s{\sigma}
\def\cG{{\cal G}}
\def\tcG{\widetilde{\cal G}}
\def\cJ{{\cal J}}
\def\cT{{\cal T}}
\def\cV{{\cal V}}
\def\cW{{\cal W}}
\def\cU{{\cal U}}
\font\ensX=msbm10
\font\ensVII=msbm7
\font\ensV=msbm5
\newfam\math
\textfont\math=\ensX \scriptfont\math=\ensVII \scriptscriptfont\math=\ensV

\title{Hybrid formalism and topological amplitudes\footnote{Contribution
    to the proceedings of the ICMP 2006 conference (August
    2006, Rio de Janeiro, Brazil). SPhT-T07/006, hep-th/0607021}}   

\author{{{\bf J\"urg K\"appeli}${}^{1}$, 
{\bf Stefan Theisen}${}^{2}$, and {\bf Pierre Vanhove}${}^{3}$}\\[1.5em]
{${}^{1}$ \it Humboldt Universit\"at, Institut f\"ur
  Physik, Berlin, Germany}\\[0.5em]
{${}^{2}$ \it Max-Planck-Institut f\"ur Gravitationsphysik,}\\
{\it Albert-Einstein-Institut, Golm, Germany}\\[0.5em]
{${}^{3}$\it   CEA/DSM/SPhT, URA au CNRS, CEA/Saclay,}\\
{\it F-91191 Gif-sur-Yvette, France}\\[1em]
{\tt kaeppeli@aei.mpg.de, theisen@aei.mpg.de, pierre.vanhove@cea.fr}}

\begin{document}

\maketitle

\abstract{We study four-dimensional compactifications of type II
superstrings on Calabi-Yau spaces in the hybrid formalism.
Chiral and twisted-chiral interactions are re\-de\-rived, which involve
the coupling of the compactification moduli to two powers of the
Weyl-tensor and of the derivative of the universal tensor field-strength.
We review the formalism and provide details of some of its technicalities.}


\section{Introduction}

Type II string compactified on a Calabi--Yau 3-fold gives rise to
${\cal N}=2$ supergravity in four dimension. Most calculations 
of string scattering amplitudes, and therefore of the construction 
of the low-energy-effective action, are based on the  
Ramond-Neveu-Schwarz (RNS) formulation 
of the superstring. A drawback of this formulation is that spacetime 
supersymmetry is not manifest and is achieved only after GSO projection. 

An alternative formulation without these complications is the hybrid
formulation.  Hybrid string theory can be obtained by a field
redefinition from the gauge-fixed RNS string or by covariantizing the
Green-Schwarz (GS) string in light-cone gauge.  In this sense,
worldsheet reparametrizations are gauge-fixed in the hybrid
formulation. Nevertheless, there is no need for ghost-like fields in
the formalism since the theory can be formulated as a ${\cal N}=4$
topological theory and amplitudes can be computed directly by the
methods of topological string theory \cite{BerkovitsVY}. The
theory consists of two completely decoupled twisted worldsheet SCFT,
one describing the spacetime part, one the internal part. Despite
being twisted, hybrid string theory describes the full theory, i.e.,
it computes also non-topological amplitudes. Hybrid type IIA and IIB
string theories are distinguished by the relative twisting of the
left- and right-moving sector of the internal SCFT. When working with
either type one is therefore committed to a given fixed twisting.

The hybrid formulation was developed in a series of papers by
N. Berkovits and various collaborators
\cite{BerkovitsVY,BerkovitsCY,BerkovitsCB}. It was reviewed in
\cite{BerkovitsBF}. One purpose of this note is our attempt to fill in
details of some of the more technical aspects.  This is done in
Sections 2 and the Appendices B and C, the content of which is well
known to the few experts in the field, but often not readily
accessible.

The main application of hybrid strings in this note are presented in
Sections 3 and 4. We extend the analysis of higher order derivative
interactions to the twisted-chiral sector. The procedure is analogous
to the computation in the chiral sector given \cite{BerkovitsVY}. Even
though one is working with a fixed relative twisting, giving rise to
either type IIA or type IIB, it is shown that the chiral and
twisted-chiral couplings of each type II theory depend on both the
A-model and B-model topological partition functions. In the effective
action these amplitudes give rise to couplings of compactification
moduli to two powers of the Weyl tensor or of the derivative of the
universal tensor field-strength. In the RNS formulation these
couplings were discussed in \cite{AGNT}.

Another possible application is flux compactifications of the type II
string with ${\cal N}=1$ spacetime supersymmetry. The breaking ${\cal
N}=2\to{\cal N}=1$ results from auxiliary fields acquiring vacuum
expectation values \cite{Vafa:2000wi}. Due to its manifest spacetime supersymmetry, the
hybrid formulation might be the most suitable.  First steps in this
direction were already taken in \cite{Lawrence:2004zk,Linch:2006ig}.

\section{Compactified string theory in RNS and hybrid variables}

In this section we present a detailed account of the field mapping
between the variables of the Ramond-Neveu-Schwarz (RNS) formulation of
those of the hybrid formulation of the superstring.  We consider here
only Calabi-Yau compactifications to four spacetime dimensions and
split all variables into a spacetime and an internal part. The
internal part is practically the same for the RNS and the hybrid
formulation, while the two descriptions of the spacetime are
different.

\subsection{Hybrid variables}
\label{sec:hyb}

Type II and heterotic string theory compactified on a Calabi-Yau three-fold can be 
formulated within a covariant version of the Green-Schwarz (GS) formulation \cite{
BerkovitsCY,BerkovitsCB}. The spacetime part consists of four bosons $x^m$, 
two pairs of left-moving canonically conjugate 
Weyl fermions $(p_\a,\t^\b)$ and $(\bar p^\ad,\tb_\bd)$, both of
conformal weight $(1,0)$ and a chiral boson $\rho$ with action
\begin{equation}\label{hybridaction}
S={1\over\pi}\int d^2z \left\lbrace{1\over2}\pb x^m\p x_m+p^\a\pb\t_\a
+\bar p_\ad\pb\tb^\ad+\tilde p^\a\p\tilde\t_\a+\bar{\tilde
p}_\ad\p\bar{\tilde\t}{}^\ad 
+\half \pb\rho\p\rho\right\rbrace\,.
\end{equation}
The chiral boson is periodic with period $\rho\sim \rho+2\pi i$ and 
\begin{equation}\label{eRhoOPE}
\rho(z) \rho(w) = -\ln(z-w)\, ,
\end{equation}
From these fields one constructs
the generators 
\begin{equation}\begin{split}\label{eGenerators}
T&= -{1\over 2} \p x^m\p x_m - p^\a\p\t_\a - \bar
p_\ad \p\tb^\ad  
-{1\over2} \partial \rho \partial\rho - \frac12 Q_\rho  \partial^2
{\rho}\,,\\
 G^-&={1\over \sqrt{32}} e^{\rho}\,d^2\,,\qquad
G^+=-{1\over \sqrt{32}}e^{-\rho}\, \bar d^2\,,\\
 J&=\partial \rho\,,
\end{split}
\end{equation}
We have defined the fermionic currents (cf. Appendix~\ref{secConvention})
\begin{equation}
\begin{split}
\label{eDefd}
d_\a&=p_\a+i\tb^{\ad}\p x_{\a\ad}-\tb^2\p\t_\a+{1\over2}\t_\a\p\tb^2\,,\\
 \bar d^{\ad}&=\bar p^\ad+i\t_\a\p
x^{\a\ad}-\t^2\p\tb^\ad+{1\over2}\tb^\ad\p\t^2\,.
\end{split}
\end{equation}
In the definition of the energy-momentum tensor a background charge
$Q_\rho$ for the chiral boson $\rho$ is included. It
is obtained from the coupling of the field $\rho$ to the world-sheet
curvature. This coupling is not visible in conformal gauge. The
background charge implies the conformal weights ${\rm
  wt}(\exp(q\rho))=-\frac12 q(q+Q_\rho)$, and therefore ${\rm
  wt}(\exp(\pm \rho))=-{1\over2}\, (1\pm Q_\rho)$ and ${\rm
  wt}({G^\pm})={1\over 2}\, (3\pm Q_\rho)$. Also the central charge of
the Virasoro algebra depends on the value of $Q_\rho$. It is
$c_x+c_{p,\theta}+c_{\bar p,\bar\theta}+c_\rho=4-4-4+(1+3 Q_\rho^2)=
3\,(Q^2_\rho-1)$. For $Q_\rho=0$, $(T,G^+,G^-,J)$ generate an
untwisted $c=-3$, ${\cal N}=2$ superconformal algebra while for
non-vanishing $Q_\rho$ the algebra is twisted. It is topological for
$Q_\rho=\pm 1$. When checking the algebra for this case, in
particular, the correct overall sign on the right-hand side of
\begin{equation}\label{GplusGminus}
G^+(z)G^-(w)\sim{{c\over3}\over(z-w)^3} +
{J(w)\over(z-w)^2} + {T(w)\over z-w}\,,
\end{equation}
for $c=-3$, the relative minus sign in the definitions of $G^\pm$ is crucial. 
It is also consistent with the requirement $(G^+)^\dagger=G^-$ if we define 
$(e^\rho)^\dagger=-e^{-\rho}$. 
The hermiticity properties of the hybrid variables are further discussed 
in Appendices~\ref{secConvention} and~\ref{appMap}. 
As explained in
section~\ref{secFR}, the field mapping from the RNS 
variables determines the background charge as $Q_\rho = -1$. The
four-dimensional part is therefore a twisted $c=-3$, ${\cal
N}=2$ superconformal algebra.

The Calabi-Yau compactification is described by an internal 
${\cal N}=2$ SCFT. The generators $(T_C, G^+_C, 
G^-_C,  J_C)$ form an untwisted 
$c=9$, ${\cal N}=2$ superconformal algebra and commute with
(\ref{eGenerators}).  The generators $({\cal
T}, {\cal G}^{\pm},{\cal J})$ of the combined system are obtained by
adding\footnote{When working with the explicit realizations
  (\ref{eGenerators}) of $G^{\pm}$ cocycle factors must be included
  in order for the space-time and the internal part of $\cG^\pm$ to
  anticommute. The explicit expressions are given in
  (\ref{eq:fermgen}).} the twisted    
generators $( T_C+{1\over2}\p  J_C,  G^+_C,  G^-_C,  J_C)$ 
to those of (\ref{eGenerators}), 
\begin{equation}\label{hybrGen}
{\cal T} = T +  T_C + {1\over2}\p  J_C\,,\quad
{\cal G}^\pm = G^{\pm}+  G^{\pm}_C\,,\quad {\cal J} = J+ J_C\,.
\end{equation}
They form a twisted $c=6$, ${\cal N}=2$ superconformal algebra.  
The current $J_C$ can be represented in terms of a free boson $H$ as 
$J_C=i\sqrt{3}H$. The generators $G_C^\pm$ can then be written in the form 
$G_C^+=e^{+{i\over\sqrt3}H}G'$ and $G_C^-=e^{-{i\over\sqrt{3}}H}\bar G'$ 
where $G'$ and $\bar G'$ are uncharged under $J_C$. The conformal weight 
of $e^{{iq\over\sqrt3}H}$ is $\frac{q}{6}(q-3)$.

For a twisted algebra, the conformal anomaly vanishes (though
the other currents are anomalous). There are, therefore, two options:
either, one untwists the resulting 
algebra, couples the system to a set of $c=-6$, ${\cal N}=2$
superconformal ghosts (thereby canceling the central charge) and
calculates scattering amplitudes utilizing the ${\cal N}=2$
prescription~\cite{Berkovits:1992bm}. Alternatively, one embeds the
twisted $c=6$, ${\cal N}=2$ SCFT into a (small 
version of the) twisted ${\cal N}=4$ algebra and uses
the topological prescription \cite{BerkovitsVY,BerkovitsIM} to
compute the spectrum and correlation functions. This is the method
we follow.

The embedding into a twisted small ${\cal N}=4$ superconformal
algebra\footnote{Small ${\cal N}=4$ superconformal algebras were constructed in \cite{Ademollo:1976wv}. 
Our conventions are based on the algebra presented in~\cite{YuRY}.} proceeds as
follows: The ${\rm U}(1)$-current ${\cal J} = J + J_C$ is augmented to a
triplet of currents $({\cal J}^{++}, {\cal J}, {\cal J}^{--})$. 
The $\cJ$-charge of
$\cJ^{\pm\pm}$ is $\pm 2$ and the conformal weights are ${\rm
  wt}({\cal J}^{++})=0$ and ${\rm wt}({\cal J}^{--})=2$.  They satisfy
the ${\rm SU}(2)$ relation
\begin{equation}
\label{eJJJ}
{\cal J}^{++}(z) {\cal J}^{--}(w) \sim {1\over(z-w)^2} + {{\cal
    J}(w)\over (z-w)}\,.   
\end{equation}
There are two ${\rm SU}(2)$ doublets of fermionic generators:
$({\cal G}^+,\widetilde {\cal G}^{-})$ and $({\cal G}^-,
\widetilde{{\cal G}}^+)$ that transform in
the ${\bf 2}$ and ${\bf 2^*}$ of ${\rm SU}(2)$,
respectively. The $\widetilde{{\cal G}}^\pm$ are defined via the operator
products 
\begin{equation}
  \label{edefGtilde}
{\cal J}^{\pm\pm}(z){\cal  G}^{\mp}(w) \sim \mp{\widetilde{{\cal
      G}}^{\pm}(w)\over z-w}\,, 
\quad {\cal J}^{\pm\pm}(z)\widetilde{{\cal G}}^{\mp}(w)  \sim
\pm{{\cal G}^{\pm}(w)\over z-w}\,.   
\end{equation}
and have ${\rm wt}(\tilde{\cal G}^+)=1$ and ${\rm wt}(\tilde {\cal
  G}^-)=2$. The other OPEs of ${\cal J}^{\pm\pm}$ with the fermionic
generators are finite. The notation $\widetilde{{\cal O}}$ refers to a
more general operator conjugation ${\cal O}\rightarrow \widetilde{\cal
  O}$, for which~(\ref{edefGtilde}) is a special case. It is explained
in Appendix~\ref{appMap}.

The nontrivial OPEs of the supercurrents are
\begin{equation}
  \label{eGJpp}
 {\cal G}^+(z) \widetilde{{\cal G}}^+(w) \sim {2 {\cal J}^{++}(w)\over (z-w)^2} +
{\partial {\cal J}^{++}(w)\over z-w}\,,\quad 
 \widetilde{{\cal G}}^-(z) {{\cal G}}^-(w) \sim {2 {\cal J}^{--}(w)\over (z-w)^2} +
{\partial {\cal J}^{--}(w)\over z-w}\,.  
\end{equation}
and
\begin{equation}
  \label{eq:eOPENfour}
{\cal G}^{+}(z){\cal G}^-(w)\sim \;{2\over(z-w)^3}
+ {{\cal J}(w)\over(z-w)^2} + {{\cal T}(w)\over z-w}\,,
  \end{equation}
and the very same OPE for $\widetilde{{\cal G}}^{+}(z)$ and
$\widetilde{{\cal G}}^{-}(w)$. 
The explicit form of the currents and super-currents is 
\begin{equation}
\label{eJJ}
{\cal J}^{\pm\pm}(z) = c_\pm {e}^{\pm \int^z\!\! {\cal J}} = 
c_\pm{e}^{\pm(\rho+ i\sqrt3  H)}\,,
\end{equation}
and
\begin{equation}
\begin{split}
  \label{eq:fermgen}
  {\cal G}^+=&-\Big(\textstyle{1\over\sqrt{32}}e^{-\rho}\bar d^2+c_+G_C^+\Big)\\
  {\cal
    G}^-=&\phantom{-\Big(}\,\,\,\textstyle{1\over\sqrt{32}}e^{\rho}d^2+c_-G_C^-\\ 
  \tilde {\cal G}^+=&-\Big(\textstyle{1\over\sqrt{32}}
  c_+ e^{2\rho+i\sqrt{3}H}d^2+e^\rho G_C^{++}\Big)\\
  \tilde {\cal G}^-=&-\Big(\textstyle{1\over\sqrt{32}} c_-
  e^{-2\rho-i\sqrt{3}H}\bar d^2+e^{-\rho} G_C^{--}\Big)
\end{split}
\end{equation}
Here ${ G}{}_C^{\pm\pm}$  
are defined\footnote{
The expression $A(B(w))$ denotes the residue in the OPE of $A(z)$ with
$B(w)$ and equals the (anti)commutator $[\oint A, B(w)\}$.  The notation is $\oint A \equiv = \tfrac{1}{2\pi i} \oint dz A(z)$.
}%
{} as ${{ G}}{}_C^{\pm\pm} = e^{\pm i\sqrt{3}H}( G^{\mp}_C)$
and $c_\pm=e^{\pm i\pi\oint {\cal J}}=e^{\pm
  i\pi(p_\rho+\sqrt{3}p_H)}$.\footnote{The momentum modes $p_\rho=\oint\p\rho$ 
and $p_H=i\oint\p H$ 
satisfy the commutation relations $[p_\rho,\rho]=-1$ and
$[p_H,H]=-i$. Their hermiticity properties are discussed in
Appendix~\ref{appMap} and imply $(c_+)^\dagger =c_-$.} 
The various signs and cocycle factors $c_\pm$ 
are necessary in order to guarantee the 
hermiticity relations $({\cal J}^{++})^\dagger={\cal J}^{--}$, $({\cal G}^+)^\dagger={\cal G}^-$
and $(\tilde{\cal G}^+)^\dagger=\tilde{\cal G}^-$ , the
appropriate Grassmann parity of the generators, and for correctly reproducing the  
algebra.

In type II theories the spacetime fields are supplemented by two pairs of
right-moving canonically conjugate Weyl fermions and a periodic
right-moving chiral boson. We will
use the subscripts ``$L$'' and ``$R$'' in order to distinguish
left-moving from right-moving fields and adopt the notation
$|A|^2=A_L A_R$. For notational simplicity we discuss mostly type IIB
string theory, for which the left- and right-movers
are twisted in the same way. For type IIA theories the 
right-moving part of the algebra is obtained by the opposite twisting
as compared to IIB. Operationally, the expressions for IIA can be obtained from
those of IIB by replacing $( J_C)_R \rightarrow - ( J_C)_R$ (thereby
reversing the background charge) in above definitions of the currents
and by reversing, e.g., $( G_{C}^{\pm})_R\rightarrow ( G_{C}^{\mp})_R$. The
spacetime part remains unaffected.

\subsection{RNS variables}

In the RNS representation the spacetime fields are $(x^m, \psi^{m})$ with
$m=1,\cdots ,4$.  They contribute with $c^{x,\psi}=6$ to the central
charge of the Virasoro algebra. We will  
concentrate on the left-moving sector in what follows.

It is convenient to bosonize the (Euclideanized) worldsheet fermions, 
\begin{equation}\label{eBoson}
\psi^{1}\pm i \psi^{2}=e^{\pm i\varphi^{1}}\,,\qquad 
\psi^{3}\pm i \psi^4=e^{\pm i\varphi^2}\,.
\end{equation}
As usual we suppress cocycle factors. 
The bosonized expression for the
${\rm SO}(4)$-spin fields of positive and negative chirality are  
\begin{equation}\label{eSpinFields}
S^{\a}=e^{\pm{ i\over2}( \varphi^1+\varphi^2)}\,,\qquad
\bar S^{\ad}=e^{ \pm {i\over2}(\varphi^1- \varphi^2)}\,.
\end{equation}
The internal
sector (the Calabi-Yau
threefold) is accounted for by a $c=9$ CFT with ${\cal N}=2$
worldsheet superconformal symmetry 
generated by $\breve T_C$, $\breve G_C{}^\pm$, and $\breve J_C$. 
Their relation to the generators introduced in
the previous section is explained in sec.~\ref{secFR}.
The ${\rm U}(1)$ R-current
$\breve J_C$ can be expressed in terms of a free chiral boson $\breve H$ as
\begin{equation}\label{eBosU}
\breve J_C =i\sqrt3 \, \partial \breve H\,,\quad \breve H(z) \breve H(w)=-\ln(z-w)\,.
\end{equation}
Any field ${\cal O}^{(q)}$ with $R$-charge $q$ can 
be decomposed as ${\cal O}^{(q)}=\exp({iq\over\sqrt{3}}\breve H){\cal O}'$ where 
${\cal O}'$ is uncharged with respect to $\breve J_C$. For the generators
$\breve G{}_C^{\pm}$ this part is independent of $\breve H$.

Covariant quantization requires fixing the local reparametrization
invariance of ${\cal N}=1$ worldsheet supergravity. This introduces
the $(b,c)$ and $(\beta,\gamma)$ ghost systems.  
With $c^{{\rm gh}}=-15$ the total central charge vanishes. Following \cite{FMS},
we `bosonize' the ghosts
\begin{equation}
\label{eBosGhost}
\begin{split}
b &= e^{-\sigma}\,,\qquad\qquad\qquad\quad\,\, c=e^\sigma\,,\\
 \b &= e^{-\phi}\p\xi = e^{-\phi+\chi}\p\chi \,,\quad \gamma=e^\phi\eta =
 e^{\phi-\chi}\,,\\
 \xi&=e^\chi\,,\qquad \qquad \qquad\qquad\eta=e^{-\chi}\,.
\end{split}
\end{equation}
The total energy-momentum tensor is
\begin{equation}
\begin{split}
\label{eTRNS}
T_{{\rm RNS}} =&  -{1\over 2} \p x^m \p x_m- {1\over 2} \psi^m\p\psi_m +
\breve T_C \\
& +  {1\over 2} \left[ (\partial\sigma)^2 +
(\partial\chi)^2 - (\partial\phi)^2\right]  - {1\over2}\, \partial^2
(2\phi -3\sigma-\chi)\,.
\end{split}
\end{equation}
The generators
$(T_{\rm RNS},\, b,\, j_{\rm BRST},\, 
J^{\rm gh})$ form a twisted ${\cal N}=2$ algebra. The ${\rm U}(1)$-current is
$J^{{\rm gh}} = -(bc + \xi \eta)$. The BRST-current
$j_{{\rm BRST}}$ is given in Appendix~\ref{appMap}.

%
%

\subsection{Field redefinition from RNS to hybrid variables}
\label{secFR}

The RNS variables are mapped to the hybrid variables in a two-step procedure.
From the RNS variables one first forms a set of variables, called the
``chiral GS-variables'' in \cite{BerkovitsCY,BerkovitsVY}. In this
section, we refer to these variables. The hybrid variables of the previous section 
are obtained in a second step by performing a field redefinition on the
chiral GS-variables. We will suppress this field redefinition in the
following and refer to Appendix~\ref{appMap}\ for a detailed account.

Following \cite{BerkovitsCY,BerkovitsVY} we define the following
superspace variables:\footnote{The RNS variables are actually subject to
the rescaling given in
Appendix~\ref{sec:fr}. We neglect this issue here.} 
\begin{equation}\label{eMapTheta}
\begin{split}
\theta^{\alpha} &=c\xi\, e^{-{3\over2}\phi} \, \bar{\Sigma}\,
S^{\alpha}\,, \qquad
\bar\theta^{\dot\alpha}= e^{\phi\over2} \, \Sigma\, \bar S^{\dot
\alpha}\,,\\
 p_{\alpha}&=  b \eta\, e^{{3\over2}\phi}\, \Sigma \, S_{\alpha}\,,
\qquad \, \,\bar p_{\dot\alpha}= e^{-{\phi\over2}}\, \bar{\Sigma}\, \bar
S_{\dot\alpha}\,,
\end{split}
\end{equation}
where 
\begin{equation}\label{Hdef}
\Sigma=e^{{i\over2}\sqrt{3}\breve H}\,,\quad
\bar{\Sigma}=e^{-{i\over2} \sqrt{3}\breve H}\,.
\end{equation}
In this definition, $\t^\a$ ($p_\a$) carries charge $q_C=-{3\over2}$
$(+{3\over2})$.  Here and in the following, $q_C$
denotes the charge under $\oint \breve J_C$, the ${\rm U}(1)$ R-symmetry of the
internal $c=9$ SCFT.

In order to implement a complete split between the spacetime and the
internal part one must require that the hybrid variables
(\ref{eMapTheta}) do not transform under the $c=9$ SCFT generators.
For instance, the variables~(\ref{eMapTheta}) should not carry a
charge with respect to the ${\rm U}(1)$ R-symmetry. This can be
realized by shifting the ${\rm U}(1)$ charge by the picture-counting
operator,
\begin{equation}\label{eP}
{\cal P} = -\b\gamma + \xi\eta = -\p\phi+ \p\chi\,.
\end{equation}
 The variable $\t^\a$, for instance, has picture 
 $-{1\over 2}$ . This motivates the following definition of the
 shifted ${\rm U}(1)$ current 
\begin{equation}\label{eNewJC}
 J_C = \breve J_C - 3 {\cal P} =\breve J_C +  3\, \partial(\phi
 -\chi)\,. 
\end{equation}
More generally, the fields of the internal part are transformed by the
 field transformation~\cite{BerkovitsCY,BerkovitsVY}
\begin{equation}\label{ePicTw}
 F_C = e^{\cal W} \breve F_C e^{-{\cal W}}\,,\quad {\cal W} =
\oint  (\phi -\chi) \breve J_C\,.
\end{equation}
For the other generators of the internal ${\cal N}=2$ algebra 
this implies 
\begin{equation}\label{eNewGenerators}
\begin{split}
 G^+_C &= e^{(\phi -\chi)} \breve{G}{}_C^+\,,\\
  G^-_C &= e^{-(\phi - \chi)} \breve G{}^-_C\,, \\
 {T}_C & = \breve T_C + \partial(\phi -\chi)\, \breve J_C+ {3\over 2}\, 
\left(\partial \phi- \partial  \chi\right)^2 \,. 
\end{split}
\end{equation}
These are the generators that couple to the chiral GS-variables
defined in (\ref{eMapTheta}).  
The generators coupling to the hybrid variables are related to these
by the field redefinition  
discussed in Appendix~\ref{appMap}, which does not affect
the algebraic structure  
discussed in the following.
The currents $( T_C, G^+_C,  G^-_C,  J_C)$ generate an
untwisted ${\cal N}=2$ superconformal algebra.
The shift by the picture changing current in the relation
(\ref{eNewJC}) amounts to a background
charge $Q_{ J_C}=-3$ for the current $ J_C$.
The RNS ghost-current 
\begin{equation}\label{eRNSghostC}
J^{\rm gh} = -(bc + \xi \eta)= \partial \sigma -\partial \chi\,,
\end{equation}
which is obtained from the ghost
current of the ``small Hilbert space'' $
-(bc+\b\gamma)=\p\sigma-\p\phi$ by adding 
the picture-counting operator (\ref{eP}), is 
mapped to a combination of 
the current $J=\partial\rho$ and the shifted internal ${\rm U}(1)$
R-current \cite{BerkovitsCY,BerkovitsVY},
\begin{equation}\label{eDefRho}
J=\p\rho= J^{\rm gh}- J_C =  \p \sigma +2\p\chi-3\p\phi - \breve
J_C\,.
\end{equation}
This equation defines the chiral boson $\rho$ in terms of the RNS
variables. The mapping is such that the $\rho$-system\footnote{The
  current which satisfies $T(z) j(w) \sim
  \frac{Q_\rho}{(z-w)^3}+\ldots$ and leads to $\bar\partial j = \tfrac{1}{8}
  Q_{\rho}\sqrt{g} R$  is $j=-\p\rho = -J$.} acquires a background
charge $Q_\rho=-1$ and that it has regular OPEs with the internal
generators $( T_C, G^+_C, G^-_C, J_C)$.  The superspace variables
$\theta$, $\bar\theta$, $p$ and $\bar p$, and the redefined internal
operators~(\ref{ePicTw}) all have zero $\rho$-charge. This, in
particular, means that (\ref{ePicTw}) leads to a complete decoupling
of the internal sector from the chiral GS-variables.

The field redefinitions (\ref{eMapTheta}) are such that the RNS generators
$(T_{\rm RNS}\,, b\,,j_{\rm BRST}\,,\\J^{\rm gh})$ map to the hybrid generators of
the ${\cal N}=2$ algebra 
\begin{equation}\label{eThTRNS} 
T_{\rm RNS} = {\cal T}\,,\quad b = {\cal
G}^- \,,\quad j_{\rm BRST} = {\cal G^+}\,, \quad J^{\rm gh} = {\cal J}\,. 
\end{equation}
We hasten to add that in order to arrive at this
correspondence one must correctly take into account the field mapping
from the chiral GS-variables to the hybrid variables
(cf. Appendix~\ref{appMap}).

It is straightforward to express the raising and lowering operators
(\ref{eJJ}) of
the ${\cal N}=4$ algebra in terms of RNS
variables, since one can verify that these are not affected by the additional field
redefinition, mapping  hybrid to chiral GS-variables as discussed in
Appendix~\ref{appMap}. From (\ref{eMapTheta}) one therefore concludes 
\begin{equation}\label{jJppJmm}
{\cal J}^{++} = c\,\eta\,,\qquad
{\cal J}^{--} =  b\,\xi\,.
\end{equation}
Using this it is easy to verify that the generators
$\widetilde{\cG}^\pm$, defined through (\ref{edefGtilde}), are
expressed in RNS variables by
\begin{equation}\label{eDic}
\widetilde{\cG}^{-} =\left[ Q_{\rm BRST}, b\,\xi\right]= b\,
Z+\xi\,T_{\rm RNS}\,,\quad \widetilde{\cG}^{+}  = \eta\,,
\end{equation}
where $Z$  is the picture changing operator of the 
RNS formalism, given in~(\ref{picchop}). 
We summarize the dictionary between the RNS and the hybrid currents
in the following table:
\begin{eqnarray}\label{DicSum}
\cT&=& T_{\rm RNS} \,,\\
\nn  \cJ^{++}=c\, \eta \,,\quad \cJ^{--}&=&  b\,\xi \,,\quad 
\cJ = J^{\rm gh} = -(bc+\xi\eta)\,, \\
\nn \cG^+=j_{\rm BRST} \,,\quad {\widetilde \cG}^{+} &=& \eta \,,
\quad\, \cG^- = b\,, \quad \widetilde{\cG}^{-} =  b\,Z+\xi T_{\rm RNS}\,.
\end{eqnarray}

So far we have concentrated on the left-moving (holomorphic) sector 
of the theory. For the heterotic string the right-moving sector 
is treated in the same way as in the RNS formulation: it is 
simply the bosonic string. For the type II string, however,
the distinction between type IIA and IIB needs to be discussed. Since the 
construction presented above involves twisting the internal $c=9$ SCFT, 
the distinction between 
IIA and IIB is analogous to the one in topological string theory where 
one deals with the so-called A and B twists (which are related by mirror symmetry). 
In type IIB, the left- and right-moving sectors are treated identically and the distinction is 
merely in the notation, i.e., to replace all fields $\phi_L(z)$ by $\phi_R(\bar z)$. 
In type IIA, however, the twists in the two sectors are opposite.
The two possible twists differ in the shift of the conformal weight, which is 
either $h\to h- {1\over2}q$ or $h\to h+{1\over2}q$. Above we have discussed the  
first possibility. The second twist is implemented by the replacement 
$\breve T_C\to \breve T_C-{1\over2}\partial \breve J_C$ and follows from the first by the 
substitution $\breve J_C\to -\breve J_C$. 
This also implies that the transformation (\ref{ePicTw}) is now defined with 
${\cal W}=-\oint(\phi-\chi)\breve J_C$ which leads to 
\begin{equation}
\begin{split}
 T_C&=\breve T_C-\p(\phi-\chi)\breve J_C+{3\over2}(\p\phi-\p\chi)^2\,,\\
{J}_{C}&= \breve J_{C} + 3 {\cal P}\,,
\end{split}
\end{equation}
and 
\begin{equation}\label{eRightJRho}
J= \partial\rho=J^{\rm gh}+ J_C
=\partial\sigma+2\p\chi-3\p\phi+ \breve J_{C}\,.
\end{equation}
With this definition, $\rho$ still has background charge
$Q_{\rho}=-1$.  The twisted $c=9$ SCFT is generated by
$({T}_{C}- {1\over 2}\partial {J}_{C},
G_{C}^{+}, G_{C}^{-}, {J}_{C})$, where now the conformal
weights of $ G^{+}_{C}$ and $ G^{-}_{C}$ are two and one,
respectively.  The full right-moving supersymmetry generators for the
type IIA theory are (suppressing signs and cocycle factors)
\begin{equation}\label{eFullRG}
{\cal G}^{\pm}_R=G^{\pm}_R+{G}{}^{\mp}_{C\, R}\,.
\end{equation}
The map between RNS and hybrid variables must also be modified for the latter to be 
neutral under ${J}_C$:
\begin{equation}\label{eRTheta}
\begin{split}
  \theta^{\alpha} &= c\xi\, e^{-{3\over2}\phi}\,{\Sigma}\,
  S^{\alpha}\,, \qquad
  \bar{\theta}^{\dot\alpha}=e^{\phi\over2}\,\bar{\Sigma}\, \bar
  {S}^{\dot\alpha}\,,\\  
  p_{\alpha}&= b \eta\, 
  e^{{3\over2}\phi}\,\bar{\Sigma} \, S_{\alpha}\,,
  \qquad \, \,\bar
  {p}_{\dot\alpha}=e^{-{\phi\over2}}\,{\Sigma}\,\bar{S}_{\dot\alpha}\,. 
\end{split}
\end{equation}
In the type IIA theory this applies for the right-movers, given
(\ref{eMapTheta})\ for the left movers.  Summarizing, the difference
between type IIA and type IIB is seen in the different right-moving
${\rm U}(1)$ charge assignment to $\rho_R$.

\subsection{Physical state conditions and ${\cal N}=4$-embeddings}
\label{secNfour}

Having the dictionary~(\ref{DicSum}) at hand it is simple to rephrase
the standard physical state conditions of the RNS formalism in terms
of hybrid variables. We refer to \cite{BerkovitsBF,BerkovitsIM} for
details. Physical RNS vertex operators are in the cohomology of
$Q_{\rm BRST}=\oint j_{\rm BRST}$ and $\oint\eta\,$:
%
\begin{equation}\label{cohoRNS}
j_{\rm BRST}(V^+) = 0\,,\quad \eta (V^+) 
 = 0\,,\quad \delta V^+ = j_{\rm BRST}(\eta(\Lambda^-))\,,
 \end{equation}
The condition imposed by $\oint \eta$ implies that $V^+$ is in the small
RNS Hilbert space, i.e., it does not depend on the $\xi$
zero-mode. Furthermore, $V^+$ has ghost number 1 with respect to
(\ref{eRNSghostC}) as indicated with the superscript. The charge with
respect to $ -(bc+\beta\gamma)$ is $1+\cal{P}$, where
${\cal P}$ is the picture (\ref{eP}). Using (\ref{DicSum}), the conditions
(\ref{cohoRNS}) are expressed in hybrid variables as 
\begin{equation}\label{haba}
\cG^+ (V^+) = 0\,,\quad 
 \tcG^{+} (V^+) = 0\,,\quad \delta V^+ = \cG^+(
\tcG^{+}(\Lambda^-))\,,
\end{equation}
In addition, $V^+$  has ${\cal J}$-charge 1 as indicated. Note that
${\cG^+}$ and $\tcG^{+} $ have trivial cohomologies, 
since
\begin{equation}\label{triv}
\cG^+\left(\sqrt2  e^{\rho} \tb^2\right) = 1\,,\quad \tcG^{+} 
\left(\sqrt2 \widetilde{(e^{\rho} \tb^2)}\right) = 1\,.
\end{equation}
Therefore, one can solve, e.g., the
$\tcG^{+}$-constraint by introducing the ${\rm U}(1)$-neutral field $\cV$
satisfying
\begin{equation}\label{hulahaba} V^+ = \tcG^{+}( \cV) \,.
\end{equation}
Up to the gauge transformations $\delta \cV = \tcG^{+} (
\widetilde{\Lambda}^{-})$, $\cV$ is determined in terms of $V^+$ by
$\cV = \sqrt2\widetilde{(e^{\rho}\t^2)} V^+$, where we used
(\ref{triv}). It follows that (\ref{haba}) can be
rephrased in terms of ${\cal V}$ as
\begin{equation}\label{ecohozz}
\cG^+ (\tcG^{+}( \cV)) = 0\,,\quad \delta \cV = \cG^+(
\Lambda^-) + \tcG^{+}(
\widetilde{\Lambda}^{-}) \,. 
\end{equation}
Using RNS variables these manipulations become much more
transparent. Notice that $\sqrt{2} \widetilde{(e^{\rho}
\tb^2)}= e^{-2\rho-i\sqrt3 H}\t^2 = \xi$. The first equality follows
from the de\-fi\-ni\-tion~(\ref{eOs}), the second from~(\ref{eMapTheta}) (the
additional conjugation in Appendix~\ref{sec:simtr} does not affect
this result). Therefore, $\cV$ lives in the large RNS Hilbert space. Using
the RNS variables it is straightforward to show that $\cG^+(\cV) = Z 
V^+ = Z\, \tcG^{+}(\cV) $, hence $\cG^+(\cV)$ and $ \tcG^{+}(\cV)$ are
related by picture changing. This will play a role momentarily
when we discuss integrated vertex operators. From now on we will often
drop the bracket on expressions like $\cG^{\pm}(\cV)$ when the
generators $\cG^{\pm}$ and alike are involved, i.e.,
$\cG^\pm\cV\equiv\cG^\pm(\cV)$. 

Following \cite{BerkovitsIM} we fix the gauge symmetry (\ref{ecohozz})
by choosing a gauge condition which resembles Siegel gauge: we
require the vanishing of the second-order poles in the OPE's of $\cG^-$ and
$\tcG^{-}$ with $\cV$. Vertex operators $\cV$ in
this gauge have conformal weight 0. For ${\rm
  SU}(2)$ singlets these gauge fixing conditions are equivalent to the
vanishing of the second-order poles of $\cG^-$ and $\cG^+$ with
$\cV$. For massless fields
$\cV(x,\t,\tb)$ that depend only on $x$, $\t^\a$, and $\tb^\ad$ but
not their derivatives, there are no poles of order 3 or higher in the
OPE of $\cV(x,\theta,\bar\theta)$ with $\cG^{\pm}$. Hence for these
operators the gauge fixing constraints are equivalent to the primarity
constraints of the ${\cal N}=2$ subalgebra, which here means the vanishing of all
poles of order 2 and higher in the OPE of $\cV$ with $\cG^\pm$. 
This has also been explained in
\cite{BerkovitsBF,BerkovitsVY,BerkovitsCB} and we will use these
gauge-fixing constraints in the next sections also for massless fields
that depend non-trivially on the compactification. So far we have
discussed the unintegrated vertex operators $V^+$ and $\cV$ residing
in the small and large Hilbert spaces, respectively. To construct
integrated vertex operators one proceeds like in the RNS formulation:
$\int b( V^+) = \int \cG^- V^+$. Note that for this choice the
integrated and the unintegrated vertex operators are in the same
picture $\cal P$. To obtain different pictures one considers
$\int\tcG^{-}V^+$. As is shown in \cite{BerkovitsIM}, this provides
the integrated vertex operator in a different ghost picture, $\int
\tcG^{-} V^+ = \int  \cG^{-}(Z_0 V^+)$. Expressing the
operators $V^+$ in terms of $\cV$ opens new though related
possibilities: using the previous result that relates $\tcG^{+}\cV $
and $\cG^{+}\cV$, one concludes that the four possible integrated
vertex operators, $\int \cG^- \tcG^{+} \cV$, $\int 
\cG^- \cG^{+} \cV$, $\int \tcG^{-} \tcG^{+}\cV$, and
$\int \tcG^{-}\cG^{+} \cV$ are all related by picture
changing.

In the next sections we will use the following canonical ghost
pictures: we take the unintegrated NS- and
R-vertex operators in the  $-1$ and $-{1\over2}$ picture,
respectively, the integrated ones in the $0$ and $+{1\over2}$
picture. Therefore, the relevant prescription is 
\begin{equation}\label{intverop}
\int  b( Z V^+) = \int \cG^- \cG^+ \cV =
\int  \tcG^{-}\tcG^{+}\cV 
\end{equation}
Adding the right-moving sector, the relevant expression for the
integrated vertex operators can be written as
\begin{equation}\label{eintVOP}
\int d^2z\,\cW(z,\bar z)=
\int d^2z \left| \cG^- \cG^+ \right|^2 \cV(z,\bar z)\,.
\end{equation}
For better readability we drop the parenthesis here and in the
following when the first-order poles in OPE with the generators
$\cG^{\pm}$ and alike are meant.

It is convenient to label the fermionic generators by indices $i, j=
1, 2$ according to
\begin{equation}\label{efla}
\cG^{+}_i = (\cG^{+}, \tcG^{+})\,,\quad \cG^{-}_i = (\cG^-,\tcG^{-})\,.
\end{equation}
They satisfy the hermiticity property $(\cG^{+}_i)^{\dagger} =
\cG^-_i$. Consider general linear combinations 
\begin{equation}\label{elcgen}
\widehat\cG^{-}_i = u_{ij} \cG^-_j\,,\quad
\widehat\cG^{+}_i = u_{ij}^* \cG^+_j\,,
\end{equation}
where the second equation follows from the first by hermitian
conjugation. Requiring that $\widehat\cG^{\pm}_i$ satisfy the same
${\cal N}=4$ relations as $\cG^{\pm}_i$ implies that $u_{ij}$ are
${\rm SU}(2)$ parameters: $u_{11} = u_{22}^* \equiv u_1$ and $u_{21}^*
=- u_{12} \equiv u_2$ with $|u_1|^2 + |u_2|^2=1$. This shows that the
${\cal N}=4$ algebra has an ${\rm SU}(2)$ automorphism group that
rotates the fermionic generators among each other. The $u_i$'s
parameterize the different embeddings of the ${\cal N}=2$ subalgebras
into the ${\cal N}=4$ algebra. More explicitly, we have
\begin{eqnarray}\label{Generatorsu}
\begin{split}
\widehat{\tcG}{}^{+} = \widehat{\cG}^+_2 & = u_1 \tilde{\cal G}^++u_2
{\cal G}^+\,,\\ 
\widehat{\cG}^- = \widehat{\cG}^-_1& = u_1 {\cal G}^--u_2\tilde{\cal G}^-\,,
\end{split}
\end{eqnarray}
and analogous expressions for $\widehat{\cG}^+=
\cG^+_1$ and $\widehat{\tcG}{}^{-}=\widehat{\cG}^-_2$, which involve
the complex 
conjugate parameters 
$u_i^*$.

It is
advantageous to formulate the physical state conditions for general
embeddings. This generalization also plays a role in the definition of scattering 
amplitudes. As will become clearer in section~\ref{Amplnote},
the choice of a specific embedding is related to working in a specific
picture in the RNS setting. Vertex operators are therefore defined in
terms of the cohomologies 
of the operators $\oint \widehat{\cG}^+$ and
$\oint\widehat{\tcG}{}^{+}$ as in~(\ref{haba}) and~(\ref{ecohozz}). Correspondingly,
integrated vertex operators have
zero total ${\rm U}(1)$-charge and can be written in the
form
\begin{equation}\label{eVO}
  {\cal U}= \int d^2z\, |\widehat \cG^{-}\widehat \cG^+|^2 {\cal V}\,.
\end{equation}
We have $\int d^2 z \, |\widehat \cG^{-}\widehat \cG^+|^2 \cV = \int d^2z\,
|\widehat{\cG}^{+}\widehat{\cG}^-|^2 \cV$ where one drops a total derivative under the
integral. Further, if $\cV$ is an ${\rm SU}(2)$-singlet one has
$\int d^2 z \,| \widehat \cG^{-}\widehat \cG^+|^2 \cV = \int d^2 z \,
|\widehat{\tcG}{}^{-}\widehat{\tcG}{}^{+}|^2
\cV$. Therefore, as will be used later,  
$\widehat{\cG}^+ {\cal U} =\widehat{\tcG}{}^{+} {\cal U} =0$. 
\subsection{Massless vertex operators}
\label{sec:maveop}
Of particular interest are the
universal, compactification independent vertex operators contained in
the real superfield $\cV=\cV(x,\theta_{L}, \bar\theta_{L},\theta_{R},
\bar\theta_{R})$ which was discussed in \cite{BS}
(cf. appendix~\ref{secMap} for some details).
It contains the degrees of freedom of ${\cal N}=2$
supergravity and those of the universal tensor
multiplet. It satisfies the ${\cal N}=2$
primarity constraints which imply transversality constraints and
linearized equations of motion for the component fields.
In the amplitude computations of the next
section, we will pick a certain
fixed term in
the $u_i$ expansion of the integrated vertex operators (\ref{eVO}),
namely 
$\int |\cG^+ \cG^-|^2 \cV = \int | \tcG^{+} \tcG^{-}|^2 \cV$. These
operators satisfy the same properties listed below (\ref{eVO}) as the
full $u_i$-dependent operators (\ref{eVO}). For this choice, the
corresponding integrated vertex operator is obtained from
the definition~(\ref{eintVOP}) and is (up to an overall
numerical factor, cf. appendix~\ref{secVOder})
\begin{equation}\label{eVopHI}
\begin{split}
\cU& =
\int d^2z\,\left| \bar d_\ad D^2\bar D^\ad-d^\a\bar D^2
D_\a\right.\\ &\hphantom{
=\int d^2z\,\left|\right.}\left. -2i\Pi_{\a\ad}[D^\a,\bar D^\ad] 
+8(\bar\Pi_\ad\bar D^\ad-\Pi^\a D_\a)\right|^2\,{\cV}(z,\bar z)\,.
\end{split}
\end{equation}
The integrated vertex operator contains (among other
parts) the field strengths 
of the supergravity and universal tensor multiplets:   
\begin{equation}\label{emu}
\int d^2 z (d_L^\a d_R^\b P_{\a\b}+d_L^\a\bar d_R^\bd Q_{\a\bd})+{\rm
h.c.}\,,
\end{equation}
where $P_{\a\b}=(\bar D^2 D_\a)_L(\bar D^2 D_\b)_R\cV$ 
and $Q_{\a\bd}=(\bar D^2 D_\a)_L(D^2\bar D_\ad)_R\cV$ are chiral
and twisted-chiral superfields\footnote{%
See appendix~\ref{sec:chirtwichir} for details.
}. 
As discussed below, on-shell, these
superfields  describe the linearized  Weyl multiplet and the
derivative of the linearized field-strength multiplet of the universal
tensor.  For later purposes we also introduce $\cU'$ and $\cU''$ defined by
$\cU= |\cG^+|^2 \cU'$ and $\cU= |\tcG^+|^2 \cU''$, i.e.,
$\cU'=\int d^2 z|e^\rho d^\a D_\a|^2\cV $ and 
$\cU''=\int d^2 z|e^{-2\rho-\int J_C}\bar d^\ad \bar D_\ad|^2 \cV$.

The complex structure moduli are in one-to-one correspondence to
elements of $H^{2,1}(CY)$ and related to primary fields $\Omega_c$ of
the chiral $(c,c)$ ring \cite{LVW}. 
The corresponding type IIB hybrid vertex
operators are\footnote{We are suppressing the indices distinguishing between
the different elements of the ring.}
\begin{equation}\label{eVcompl}
\cV_{cc} = |e^{\rho}\bar\theta^2|^2 M_c \Omega_c\,,\quad
\cV_{aa} =({\cV}_{cc})^{\dagger} = |e^{-\rho}\theta^2|^2 \bar M_c \bar
\Omega_c\,,
\end{equation} 
where $M_c$ is a real chiral superfield (vector multiplet). Note that
in the (twisted) type IIB theory $\Omega_c$ has conformal weight
$h_L=h_R=0$, while $\bar\Omega_c$ has conformal weight $h_L=h_R = 1$.
The complexified K\"ahler moduli are in one-to-one correspondence to
elements of $H^{1,1}(CY)$ and related to primary fields
$\Omega_{tc}$ of the twisted-chiral ring $(c,a)$:
\begin{equation}\label{eVKaehl}
\cV_{ca} = e^{\rho_L-\rho_R}\, \bar\theta_L^2\, \theta_R^2 \,
M_{tc} \Omega_{tc}\,,\quad \cV_{ac} =({\cV}_{ca})^{\dagger} =
e^{-\rho_L+\rho_R}\,\theta_L^2 \, \bar\theta_R^2 \,\bar M_{tc} \bar
\Omega_{tc}\,, 
\end{equation}
where $M_{tc}$ are real twisted-chiral superfields (tensor multiplets).
The conformal weight of $\Omega_{tc}$ is $h_L=0$ and $h_R = 1$. The
integrated vertex operators are
\begin{equation}\label{eVintegr}
\cU_{cc} = \int d^2 z\, M_c |G_C^-|^2 \Omega_{c} +
\ldots\,,\quad \cU_{ca} = \int d^2 z\, M_{tc} (G_{C}^-)_L (G_{C}^+)_R
\Omega_{tc} + 
\ldots \,,
\end{equation}
where we have suppressed terms involving derivatives acting on $M_c$
and $M_{tc}$. These terms carry nonzero $\rho$-charge and 
will not play a role in the discussion of the amplitudes in section~\ref{secTop}.

The vertex operators of IIA associated to elements of the $(c,c)$
(complex structure) and $(c,a)$ ring (K\"ahler) are

\begin{equation}\label{eVOIIA}
V_{cc} = e^{\rho_L-\rho_R}\, \bar\theta_L^2\, \theta_R^2 \,
M_{tc} \Omega_c\,,\quad V_{ca} = |e^{\rho}\bar\theta^2|^2 M_c \Omega_{tc}\,.
\end{equation}

For type IIA the conformal weights of $\Omega_c$ are $h_L = 0$ and
$h_R=1$ while $\Omega_{tc}$ has weight $h_L=h_R=0$. The integrated
vertex operators involve 

\begin{equation}\label{eVintegrIIA}
U_{cc} = \int d^2 z\, M_{tc} (G_C^-)_L (G_C^-)_R \Omega_{c} +
\ldots\,,\quad U_{ca} = \int d^2 z\, M_{c} (G_{C}^-)_L (G_{C}^+)_R
\Omega_{tc} + 
\ldots \,.
\end{equation}


\section{Amplitudes and correlation functions}\label{Amplnote}

In this section we review the definition of scattering amplitudes on
Riemann surfaces $\Sigma_g$ with genus $g\geq2$ as given in \cite{BerkovitsVY}. We
also collect correlation functions for chiral bosons.

\subsection{Amplitudes}

Scattering amplitudes of hybrid string theory are defined    
in \cite{BerkovitsVY} for
$g\geq 2$ as\footnote{This differs by the factor $(\det({\rm
Im}\tau))^{-1}$ from the  
expression given in \cite{BerkovitsVY} and \cite{BerkovitsIM}. We will comment on this below.} 
\begin{equation}\label{Fg}
F_g(u_{L}, u_{R})\!=\!\int_{{\cal M}}\!\! {[dm_{g}]\over\det({\rm Im}\tau)}
\prod_{i=1}^g\Big\langle\!\int d^2 v_i\prod_{j=1}^{g-1}|\
\widehat{\tcG}{}^{+}(v_j)|^2 
|\cJ(v_g)|^2\prod_{k=1}^{3g-3}|(\mu_k,\widehat
\cG^-)|^2\prod_{l=1}^{N}\cU_l\Big\rangle\,.
\end{equation}
Since $F_g(u_{L}, u_{R})$ is a homogeneous polynomial in both
$u_{iL}$ and $u_{iR}$ of degree
$4g-4$ (we are taking $U$ to carry no $u_{iL,R}$ dependence as is explained in
section~\ref{secNfour}) this definitions provides a whole set of
amplitudes $F_g^{n,m}$ given by the coefficients in the
$u_{i\,L,R}$-expansion: 
\begin{equation}\label{eAexp}
F_g(u_{L}, u_{R})\!=\!\! \sum_{n,m}\! {4g-4\choose2g-2-n}\!
{4g-4\choose2g-2-m}\! F_g^{n,m} u_{1L}^{2g-2+n}
u_{2L}^{2g-2-n} u_{1R}^{2g-2+m} u_{2\,R}^{2g-2-m}\,,
\end{equation}
where $2-2g\leq m,n\leq 2g-2$. We focus on either
the left- or right-moving sector in the following. In view of (\ref{Generatorsu})
 it is clear that $F^{n}_g$ involves $2g-2+n$ insertions
of $\tcG^{+}$ and $\cG^-$ and $2g -2 -n$ insertions of $\cG^+$ and
$\tcG^{+}$. It is shown in \cite{BerkovitsVY} that up to contact terms 
all distributions of $\tcG^{+}$'s, $\cG^-$'s, $\cG^+$'s, and
$\tcG^{+}$'s satisfying these constraints are equivalent. We can
therefore determine $F_g^{n,m}$
(\ref{eAexp}) by evaluating a single amplitude with an admissible
distribution of insertions.

In addition there is a selection rule that relies on the cancellation of
the $R$-parity anomaly \cite{BerkovitsVY}. The $R$-charge is
\begin{equation}\label{eRp}
 R = \oint\left( \partial \rho + {1\over2}\theta^\a d_\a-{1\over
2}\bar\theta_\ad 
\bar d^\ad\right)\,,
\end{equation}
with background charge $1-g$. In the RNS formulation $R$ 
coincides with the superconformal ghost-number (picture) operator,
i.e., $R=\oint {\cal P}$.
$\tcG^{\pm}$ carry $R$-charges $\mp 1$ while those of $\cG^{\pm}$ are
zero. The contribution to the $R$-charge of the insertions is $g-1-n$.
The anomaly is therefore canceled only if the vertex operators
insertions have total $R$-charge $n$. Put differently: given vertex
operators $\prod^N_{i=1} \cU_i$ with total $R$-charge $n$, the only
non-vanishing contribution to (\ref{eAexp}) is $F^n_g$. This selection
rule is completely analogous to the one that relies on picture charge
in the RNS formulation.

It is convenient to rewrite (\ref{Fg}) in the form
\begin{eqnarray}\label{Fgtwo}
&&F_g(u_L, u_R) =\\
\nn &= &\!\!\!  \int_{{\cal M}}\!\! {[dm_{g}]\over\det({\rm Im}\tau)}
\prod_{i=1}^g\Big\langle\int d^2 v_i\prod_{j=1}^{g}|\
\widehat{\tcG}{}^{+}(v_j)|^2 
\prod_{k=1}^{3g-4}|(\mu_k,\widehat
\cG^-)|^2|(\mu_{3g-3},\cJ^{--})|^2\prod_{l=1}^{N}\cU_l\Big\rangle\,.
\end{eqnarray}
This is obtained from (\ref{Fg}) by contour deformation using
$\widehat{\cG}{}^- = \oint \widehat{\tcG}{}^{+} \cJ^{--}$ and
$\widehat{\tcG}{}^{+} = - \oint \widehat{\tcG}{}^{+} \cJ$ and the fact
that $\oint\widehat{\tcG}{}^{+}$ has no non-trivial OPE with any of
the other insertions except a simple pole with $\cJ$.  Consider the
integrand of (\ref{Fgtwo}). As a function of, say, $v_1$, it has a
pole only at the insertion point of $\cJ^{--}$. But the residue
$\langle\prod_{i=2}^{g}\widehat{\tcG}{}^{+}(v_i)
\prod_{j=1}^{3g-3}(\mu_j,\cG^-)\prod_l \cU_l\rangle$ vanishes: each of
the remaining $\widehat{\tcG}{}^{+}(v_i)$ can be written as
$-\oint\widehat{\tcG}{}^{+} \cJ(v_i)$ and $\widehat{\tcG}{}^{+}$ has no
singular OPE with any of the other insertions.  Analyticity and the
fact that $\widehat{\tcG}{}^{+}$ are Grassmann odd and of weight one,
fixes the $v$-dependence of the integrand as $\det(\omega_i(v_j))$.
The $\omega_i$ are the $g$ holomorphic one-forms on $\Sigma_g$. In
(\ref{Fgtwo}) we can thus replace
\begin{equation}\label{relation}
\prod\widehat{\tcG}{}^{+}(v_i)=\det(\omega_i(v_j))\,{\prod\widehat{\tcG}{}^{+}(\tilde  v_l)\over 
\det(\omega_k(\tilde v_l))}\,,
\end{equation}
where $\tilde v_k$ are $g$ arbitrary points on $\Sigma_g$ that can be
chosen for convenience.   
Combining left- and right-movers the $v$-integrations can be performed
with the result 
\begin{equation}\label{eImtau}
\prod_{i=1}^g\int d^2 v_i |\det(\omega_k(v_l))|^2\propto \det({\rm
Im}\,\tau)\,. 
\end{equation}
$\tau$ is the period matrix of $\Sigma_g$.
Using similar arguments one can rewrite 
\begin{equation}\label{rewrite}
{1\over\det({\rm Im}\tau)}\Bigl({\int_{\Sigma_g}}|\widehat{\tcG}{}^{+}|^2\Bigr)^g 
\propto \Big|\prod_{i=1}^g{\oint_{a_i}}\widehat{\tcG}{}^{+}\Big|^2\,.
\end{equation}
The reason for the insertion $\oint\widehat{\tcG}{}^{+}$ on
every $a$-cycle of $\Sigma_g$  
was presented in \cite{BerkovitsVY,BerkovitsIM}: it projects to the reduced Hilbert
space formed by the physical fields of an ${\cal
N}=2$ twisted theory. Amplitudes for these states can be calculated
using the rules of ${\cal N}=2$ topological strings.

\subsection{Correlation functions of chiral bosons}
\label{secRho}

In this section we provide the correlation functions which are
necessary to compute the amplitudes,
cf. \cite{Verlindes,EguchiUI,Watamura}. In the hybrid formulation
there is no sum over spin structures and no need for a GSO
projection. The correlation functions are with periodic boundary
conditions around all homology cycles of the Riemann surface
$\Sigma_g$.

We start with the correlators of the chiral boson $H$:
\begin{equation}\label{eHcorrelators}
\Big\langle\prod_k e^{i{q_k\over\sqrt{3}} H(z_k)}\Big\rangle
=Z_1^{-1/2}F({\ts{1\over\sqrt{3}}}{\ts \sum q_k z_k}-Q_H\Delta)\prod_{i<j}
E(z_i,z_j)^{{1\over3}q_i q_j}\prod_l\sigma(z_l)^{{1\over \sqrt3} Q_H q_l}\,,
\end{equation}
where $Z_1$ is the chiral determinant of \cite{Verlindes}. The prime forms
$E(z,w)$ express the pole and zero structure of the correlation
function while the $\sigma$'s express the coupling to the background
charge. Of the remaining part $F$, which is due to the zero-modes of
$H$, only the combination in which the insertion points enter will be
relevant.  It is, in fact, an appropriately defined theta-function
\cite{AGNT}. Also $F(-z)=F(z)$.  In the above expression (and below), $z$
either means a point on $\Sigma_g$ or its image under the Jacobi map,
i.e., $\vec I(z)=\int_{p_0}^z\vec\omega$, depending on the context.

The $\rho$-correlation functions are subtle. The field $\rho$ is very
much like the chiral boson $\phi$ which appears in the `bosonization'
of the superconformal $(\beta,\gamma)$ ghost system in the RNS
formulation, the only difference being the value of its background
charge.  In the RNS superconformal ghost system $\phi$ is accompanied
by a fermionic spin $1$ $(\eta,\xi)$ system. Expressions for
correlation functions of products of $e^{q_i\phi(z_i)}$ which are used
in RNS amplitude calculations are always done in the context of the
complete $(\beta,\gamma)$ ghost system. Following \cite{BerkovitsVY} our
strategy will be to combine an auxiliary fermionic spin $1$
$(\eta,\xi)$ system with the $\rho$-scalar to build a bona-fide spin
$1$ $(\beta,\gamma)$ system. We then compute correlation functions as
in the RNS formulation, which we divide by the contribution of the
auxiliary $(\eta,\xi)$-system.
Following \cite{Watamura}, we obtain  
\begin{equation}\label{eRhoCorrelator}
\Bigl\langle\prod_k e^{q_k\rho(z_k)}\Bigr\rangle_{(\beta,\gamma)}=
{Z_1^{1/2}\over\theta(\sum q_k z_k-Q_\rho\Delta)}
\prod_{k<l}E(z_k,z_l)^{-q_k q_l}\prod_r\sigma(z_r)^{-Q_\rho q_r}
\end{equation}
with $Q_\rho = -1$. As in \cite{Watamura}, the correlation function had to
be regularized due to the fact that the zero-mode contribution of the
$\rho$-field diverges. The regularization involved a projection of the
$\rho$-momentum plus the momentum of the regulating $(\eta,\xi)$
system in the loops to arbitrary but fixed values. These projections
were accompanied by factors $\oint_{a_i}\eta$ for each $a$-cycle on
$\Sigma$ and one factor of $\xi$ to absorb its (constant) zero mode.
The contribution of $(\eta,\xi)$ has to be divided out in order to
obtain the regulated correlators of the $\rho$-system. This means that
(\ref{eRhoCorrelator}) must be divided by
\begin{equation}\label{etaxi}
\Bigl\langle \prod_{i=1}^g\oint_{a_i}{dz_i\over 2\pi
i}\eta(z_i)\,\xi(w)\Bigr\rangle=Z_1\,.
\end{equation}
Altogether we thus find 
\begin{equation}\label{rhofinal}
\Bigl\langle\prod_k e^{q_k\rho(z_k)}\Bigr\rangle_{\rm reg.}=
{Z_1^{-1/2}\over\theta(\sum q_k z_k+\Delta)}
\prod_{k<l}E(z_k,z_l)^{-q_k q_l}\prod_r\sigma(z_r)^{q_r}\,.
\end{equation}
A useful identity is the  `bosonization formula' \cite{Verlindes}: 
\begin{equation}\label{eIdentity}
\prod_{i=1}^g
E(z_i,w)\sigma(w)={\prod_{i<j}E(z_i,z_{j})\prod_{i=1}^g\sigma(z_i)\over  
Z_1^{3/2}\det(\omega_i(z_j))}\theta(\st_{i=1}^g z_i-w-\Delta)\,.
\end{equation}
Using this identity one finds
\begin{equation}\label{eFF}
\Big\langle \prod_{k=1}^{ g} \, e^{-\rho(z_{k})}\,
e^{\rho(w)}\Big\rangle_{\rm reg.}= {1\over Z_{1}^2\,
\det(\omega_{k}(z_{l}))}\,, 
\end{equation}
which differs by a factor of $\det({\rm Im}\tau)$ from the
corresponding expression used in \cite{BerkovitsVY}.

\section{Topological Amplitudes}\label{secTop}

\subsection{Generalities}

The expressions for $F_g^n$ that one obtains by inserting the
generators (\ref{Generatorsu}) into (\ref{Fg}) in general are very
involved. Certain restrictions are imposed by background charge
cancellation. Since the total ${\rm U}(1)$ charge of the vertex
operators is zero the insertions of $\widehat{\tcG}{}^{+}$ and
$\widehat{\cG}^-$ in (\ref{Fg}) are precisely such that they cancel the
anomaly of the total ${\rm U}(1)$ current. It is therefore sufficient
to study the constraints imposed by requiring cancellation of the
background charge of the $\rho$-field.\footnote{Since the $J_C$ current is
a linear combination of the $\p\rho$ and the total ${\rm
U}(1)$-current, background charge cancellation for $H$ is then
automatic.}  A consequence of this constraint is that if the vertex
operators are not charged under $\p\rho$ then $|n|\leq g-1$. For
$|n|<g-1$ there are several possibilities how the various parts of the
operators (\ref{Generatorsu}) can contribute.  For $|n|=g-1$ and uncharged
vertex operators there is only a single amplitude that must be
considered. These cases are studied in the following. We restrict to
the case with $2g$ vertex operator insertions. There are then just
enough insertions of $\theta$ and $p$ to absorb their zero modes an
no nontrivial contractions occur.

\subsection{$R$-charge $(g-1,g-1)$}

This amplitude was computed in the RNS formalism in \cite{AGNT}. In this
section we review the computation in the hybrid formalism of
\cite{BerkovitsVY}. Imposing $\rho$ and $H$ 
background charge saturation (\ref{Fg}) leads to\footnote{Here and in
  the following we drop certain numerical factors and use the notation
as explained below (\ref{Generatorsu}).} 
\begin{eqnarray}\label{Agc}
{\cal A}_g &=&\int_{{\cal
M}}\!\![dm_{g}]{1\over |\det(\omega_i(\tilde v_j))|^2} 
\Bigl\langle\Big|\!\!
\prod_{ j=1}^{ m}e^\rho {G}^{++}_C(\tilde v_j)\!\!\!
\prod_{ j=m+1}^{ g}\!\!\!e^{-\rho}\bar d^2(\tilde
v_j)\\
\nn &&\qquad\qquad\qquad\times 
\prod_{ l=1}^{ m}(\mu_l,e^{-2\rho-\!\int\!\! J_C}\bar d^2)\!\!\!
\prod_{ l=m+1}^{ 3g-3}\!\!\!(\mu_{l},G_C^-)\Big|^2
\cU' \cU^{2g-1}\Bigr\rangle\,.
\end{eqnarray}
We have used the fact that $\oint e^{-\rho}\bar d^2$, when pulled off
from $\cU'$,  
only gets stuck at $J(v_g)$.  
$0\leq m \leq g-1$ parametrizes different ways to saturate the
background charges.\footnote{For notational simplicity we have chosen the
same $m$ for the left- and  
for the right-movers.}
We now use the freedom to choose $\tilde v_l=z_l$ for $l=1,\dots,g$
where $z_l$ are the arguments of the Beltrami differentials $\mu_l$
(which are integrated over). This is possible since the OPEs which one
encounters are the naive products (no poles or zeros). This gives
\begin{eqnarray}\label{Agcc}
&& {\cal A}_g=\\
&=\!\!\!\!&\nn\int_{{\cal M}}\!\! [dm_{g}]\!\!\int\prod_{ l=1}^{ g} d^2 z_l
{1\over|\det(\omega_i(z_l))|^2}
\Bigl\langle\big|(\mu_l, e^{-\rho}G_C^-\bar d^2(z_l))\big|^2
\prod_{ k=1}^{ 2g-3}\big|(\mu_k,G_C^-)\big|^2 \cU' \cU^{2g-1}\Bigr\rangle
\end{eqnarray}
which is independent of $m$.\footnote{This shows that for this amplitude
all admissible distributions of vertex operators parametrized by $m$
indeed lead to the same result and that the only subtleties that
arise from contact terms are the ones analyzed in
\cite{BCOV,OoguriCP}. We are not 
aware of an argument that this is generally the case.} 
Its evaluation is straightforward. 
One easily sees that there are just enough operator insertions to absorb 
the $p$ and $\bar p$ zero modes. $\theta$ and $\bar\theta$ then also only 
contribute with their (constant) zero modes.  
The $p$ zero modes must come from the explicit $d$-dependence of the
vertex operator. The  
$(p,\theta)_L$ and $(p,\theta)_R$ correlation functions contribute a factor 
$|Z_1|^4(\det {\rm Im}\tau)^2$, where the integrals over the insertion points
have already been performed. What is left is the integral over the 
$\theta$ zero-modes which are the Grassmann odd co-ordinates of
${\cal N}=2$ chiral superspace. The spinor indices
arrange themselves  
to produce $(P_{\a\b}P^{\a\b})^{g-1}P_{\gamma\delta}D_L^\gamma D_R^\delta V$. 
The $(\bar p,\bar\theta)$ correlators give a term
$|Z_1|^4|\det\omega_i(z_l)|^4$,  
leaving only the $\bar\theta$ zero-mode integrations. They can be
performed using  
$\int (d^2\bar\theta)_L(d^2\bar\theta)_R \Psi=\bar D^2_L \bar D^2_R
\Psi|_{\bar\theta_L=\bar\theta_R=0}$. 
Since $\bar D_\ad P_{\b\gamma}=0$, the only effect of this is to convert 
$D_L^\a D_R^\b V$ to $P^{\a\b}$. 
Finally, the $\rho$-correlator gives, using (\ref{rhofinal}) and (\ref{eIdentity}), 
$(|Z_1|^4|\det \omega_i(z_l)|^2)^{-1}$. The partition function of the
$x^m$ contributes a factor $|Z_1|^{-4} (\det{\rm Im}\tau)^{-2}$. To the
given order of spacetime derivatives, the $x^m$-dependence of the
vertex operators is only through its zero mode. Combining arguments we
obtain 
\begin{equation}\label{Agfin}
{\cal A}_g=\int (d^2\theta)_L(d^2\theta)_R (P_{\a\b}P^{\a\b})^g
\int_{{\cal M}}\!\! [dm_{g}]\Bigl\langle\prod_{ i=1}^{
3g-3}|(\mu_i,G_C^-)|^2\Bigr\rangle\,.
\end{equation}
The last part of this expression is the string
partition function of the topological $B$-model:
\begin{equation}\label{eABmod}
F_g^B=\int_{{\cal M}}\!\![dm_{g}]\Bigl\langle\prod_{ i=1}^{
3g-3}|(\mu_i,G_C^-)|^2\Bigr\rangle\,.
\end{equation}
To determine the dependence of $F_g^B$ on the chiral or twisted-chiral
moduli one inserts the appropriate expressions (\ref{eVOPKaehler})  into
these correlation functions. It can be shown, using
the arguments of \cite{BCOV}, that $F_g^B$ does not depend on
perturbations induced by either $(c,a)$ or $(a,c)$ operators. It
therefore depends only on the complex structure moduli and the
amplitudes calculated are therefore vector multiplet couplings (type
IIB).

\subsection{$R$-charge $(1-g,1-g)$}\label{sec43}

Starting from (\ref{Fg}) and imposing $\rho$ and $H$-background charge saturation,
one obtains, in close analogy to (\ref{Agc}), 
\begin{eqnarray}\label{Agantichiralm}
\nn {\cal A}_g'&=&\int_{{\cal M}}\!\! [dm_{g}]{1\over |\det(\omega_i(\tilde
v_j))|^2} 
\Bigl\langle\Big|\!\!
\prod_{ j=1}^{ m}G_C^+(\tilde v_j)\!\!\!
\prod_{ j=m+1}^{ g}\!\!\!e^{2\rho+\int J_C}d^2(\tilde v_j)
\prod_{ l=1}^{ m}(\mu_l,e^{\rho}d^2)\!\!\!\\
&&\qquad\qquad\qquad\times
\prod_{ l=m+1}^{ 3g-3}\!\!\!(\mu_{l},e^{-\rho}{G}_C^{--})\Big|^2
\cU'' \cU^{2g-1}\Bigr\rangle\,.
\end{eqnarray}
$0\leq m\leq g-1$ parametrizes the different ways of saturating the
background charges. By appropriate choices of the $\tilde v_j$ this
amplitude can be  
brought to the form 
\begin{eqnarray}
\nonumber {\cal A}_g'&=&\int_{{\cal M}}\!\! [dm_{g}]\int\prod_{ j=1}^{ g}d^2 z_j
{1\over|\det\omega_i(z_j)|^2}\bigl\langle\big|
(\mu(z_j),e^\rho d^2 G_C^+(z_j))\\
&\times&\prod_{ k=1}^{
2g-3}(\mu_k,e^{-\rho} {G}_C^{--}) 
\big|^2 \cU'' \cU^{2g-1}\Bigr\rangle\,,
\label{Aganti}\end{eqnarray}
which shows that also this amplitude is independent of $m$. 
However, its evaluation is most easily done for a different choice
of the insertion points $\tilde v_j$. To fix them, we start from
(\ref{Agantichiralm})
with the choice $m=0$ and compute the $\rho$ and the $H$ correlators. 
Their product is, using (\ref{eHcorrelators}) and (\ref{rhofinal}), 
\begin{equation}\label{rhotimesH}
\begin{split}
 {1\over Z_1}&\times{F(\sqrt{3}\sum\tilde v_j-{2\over\sqrt3}\sum
z_k-\sqrt{3}w+\sqrt{3}\Delta) 
\over\theta(2\sum \tilde v_j-\sum z_k-2w+\Delta)}\\
&\times{\prod_{k< l}E(z_k,z_l)^{1\over3}\prod_j E(\tilde
v_j,w)\prod_k\sigma(z_k)\sigma(w)\over 
\prod_{i<j}E(\tilde v_i,\tilde v_j)\prod_j\sigma(\tilde v_j)}\,,
\end{split}
\end{equation}
where we have only displayed the holomorphic part. 
With the help of the identity (\ref{eIdentity}) this is equal to
\begin{equation}\label{afterI}
\begin{split}
{1\over (Z_1)^{5\over2}}&\times{F({\ts \sqrt{3}\sum\tilde
v_j-{2\over\sqrt{3}}\sum 
z_k-\sqrt{3}w+\sqrt{3}\Delta})\over \theta(2\sum\tilde v_j-\sum
z_k-2w+\Delta)} \cr
&\times{\theta(\sum\tilde v_j-w-\Delta)\over\det w_i(\tilde v_j)}
\cdot\prod_{k< l}E(z_k,z_l)^{1\over3}\prod_k\sigma(z_k)\,. 
\end{split}\end{equation}
We now choose the $g$ positions $\tilde v_j$ such that 
$\vec I(\sum\tilde v_j-w-\Delta)=\vec I(2\sum\tilde v_j-\sum
z_k-2w+\Delta)$. Then the theta functions cancel and the remaining
terms are 
\begin{equation}\label{rem}
{1\over (Z_1)^{5\over2}{\rm det} \omega_i(\tilde v_j)}\cdot
F({\ts{1\over\sqrt3}\sum z_k-\sqrt3\Delta})\cdot
\prod_{k<l}E(z_k,z_l)^{1\over3}\prod_k\sigma(z_k)\,. 
\end{equation}
This can be written as 
\begin{equation}\label{remaining}
{1\over Z_1^2\det \omega_i(\tilde v_j)}\Big\langle\prod_{ k=1}^{ 3g-3}
e^{-{i\over\sqrt{3}}H(z_k)}\Big\rangle\,.
\end{equation}
The $p,\theta,\bar p$ and $\bar\theta$ correlators are as in the
previous amplitude 
(with the roles on barred and unbarred variables interchanged) and one
finally obtains  
\begin{equation}\label{Agfinb}
{\cal A}_g'=\int (d^2\bar\theta)_L(d^2\bar\theta)_R (\bar P_{\ad\bd}
\bar P^{\ad\bd})^g 
\int_{{\cal M}}[dm_{g}]\Big\langle{\prod_{ i=1}^{
3g-3}}|(\mu_i,\check G_C^-)|^2\Big\rangle\,.
\end{equation} 
Here $\check G_C^-=e^{-{i\over\sqrt{3}}H}G'_C$
where $G'_C$ is defined to be $G_C^+=e^{{i\over\sqrt{3}}H} G_C'$.
Note that $G_C^-$ and $\check G_C^-$ both have conformal weight
two. The internal amplitude multiplying the spacetime part is the
complex conjugate of the B-model amplitude (\ref{eABmod}):  
this follows from the fact that the expression (\ref{rem}) can be written as
\begin{equation}\label{remmm}
{1\over Z_1^2\det \omega_i(\tilde v_j)}\Big\langle
{\prod_{ k=1}^{ 3g-3}}
e^{{i\over\sqrt{3}}H(z_k)}\Big\rangle_{Q_H=\sqrt3}\,,
\end{equation}
where we used (\ref{eHcorrelators})
but with the reversed background charge as compared to
(\ref{remaining}). This happens if one chooses the opposite twisting in
(\ref{hybrGen}). Since the operators $\check{G}_C^-$ and $G_C^+$
both contain the same operator $G_C'$, the internal part of the
amplitude (\ref{Agfinb}) is equal to 
\begin{equation}\label{ean}
\Big\langle{\prod_{ i=1}^{ 3g-3}}|(\mu_i,\check
G_C^-)|^2\Big\rangle_{++} = 
\Big\langle{\prod_{ i=1}^{
3g-3}}|(\mu_i,G_C^+)|^2\Big\rangle_{--}\,.
\end{equation}
The subscripts refer to the two possible twistings $T_C\to T_C+{1\over
2}\partial J_C$ and $T_C\to T_C-{1\over 2}\partial J_C$ for left- and 
right-movers. Finally, since for unitary
theories $(G_C^-)^\dagger = G_C^+$, the right-hand side of (\ref{ean}) is the
complex conjugate of $F_g^B$ given in (\ref{eABmod}), and therefore ${\cal
A}_g'$ defined in (\ref{Agantichiralm}) is the complex conjugate of the
chiral amplitude ${\cal A}_g$ of (\ref{Agc}).  

\subsection{R-charges $(g-1,1-g)$ and  $(1-g,g-1)$}

The `mixed' amplitudes with $R$-charges $(g-1,1-g)$ and 
$(1-g,g-1)$ can now be written down immediately. They are expressed 
as integrals over twisted chiral superspace and involve the 
superfields $Q_{\a\bd}$ and $\bar Q_{\ad\b}$. They are 
\begin{equation}\label{tc}
{\cal A}_g''=\int (d^2\theta)_L(d^2\bar\theta)_R (Q_{\a\bd}Q^{\a\bd})^g
\int_{{\cal M}}[dm_{g}]\Big\langle
{\prod_{ i=1}^{ 3g-3}}(\mu_i,G_C^-)_L(\bar\mu_i,\check
G_C^-)_R\Big\rangle\,+\,{\rm c.c.} 
\end{equation}
By the same arguments as given before, one shows that this type IIB
string amplitude only depends 
on deformations in the $(a,c)$ (and $(c,a)$ for the complex conjugate
piece) ring,  
i.e., on K\"ahler moduli. In type IIB, these are in tensor
multiplets. From the discussion in section~\ref{sec43} it also follows that 
\begin{equation}\label{eam}
\int_{{\cal M}}\!\![dm_{g}]
\Big\langle{\prod_{ i=1}^{
3g-3}}(\mu_i,G_C^-)_L(\bar\mu_i,\check G_C^-)_R\Big\rangle_{++}\!\!\!
=\!\! 
\int_{{\cal M}}[dm_{g}]
\Big\langle{\prod_{ i=1}^{
3g-3}}(\mu_i,G_C^-)_L(\bar\mu_i,G_C^+)_R\Big\rangle_{+-}\!\!\!\! =
F_g^A\,,
\end{equation}
which is the topological $A$-model amplitude.

So far we have computed amplitudes of type IIB string theory. 
To compute type IIA amplitudes we need to twist the left- and
right-moving internal SCFTs oppositely. In the amplitudes this induces
the following changes:  
$(G^-_C)_R\to(G^+_C)_R$ and 
$(\check G_C^-)_R\to(\check G_C^+)_R$ where $\check
G_C^+=e^{{i\over\sqrt{3}}H}\bar G_C'$. 
Due to the opposite twist, 
the conformal weights are preserved under this operation. 
For instance, the spacetime part of (\ref{Agfin}) gets combined with
$F_g^A$, that of (\ref{tc}) with $F_g^B$. According to (\ref{eVOPKaehler}) and (\ref{eVOPOneTwo}),
$F_g^A$ depends on the moduli contained in vector multiplets, $F_g^B$
on those contained in tensor multiplets.

\subsection{Summary of the amplitude computation}

We have recomputed certain chiral and twisted-chiral
couplings that involve $g$ powers of $P^2$ or $Q^2$, respectively,
using hybrid string theory. The
amplitudes involve the topological string partition
functions $F_g^A$ and $F_g^B$. $F_g^A$ depends on the moduli
parametrizing the $(c,a)$ ring, $F_g^B$ on those of the $(c,c)$
ring. 
 In type IIA or type IIB, these are contained
in spacetime chiral (vector) or twisted-chiral (tensor)
multiplets, as summarized in the table. 
\begin{table}[h]
  \centering
  \begin{tabular}[]{ll|ll}
    \multicolumn{2}{c|}{ type IIA } & \multicolumn{2}{c}{type IIB}\\[0.3em]\hline
\parbox[h]{0pt}{\vspace{2em}}$(P^2)^g F_g^A$& $(c,a)$: vector & $(P^2)^g
F_g^B$& $(c,c)$: vector\\[0.5em]
$(Q^2)^g F_g^B$& $(c,c)$: tensor & $(Q^2)^g
F_g^A$& $(c,a)$: tensor\\
  \end{tabular}
\end{table}
The dependence on the moduli
of the complex conjugate rings is only through the holomorphic anomaly
\cite{BCOV}. 
As discussed in \cite{BS,BerkovitsNV}, on-shell, the superfield
$P_{\a\b}$ describes the linearization of the Weyl multiplet. Its
lowest component is the selfdual part of the graviphoton field
strength, $P_{\a\b}|=F_{\a\b}$. The $\theta_L\theta_R$-component is
the selfdual part $C_{\a\b\gamma\delta}$ of the Weyl tensor. The
bosonic components of $Q_{\a\bd}$ are $Q_{\a\bd}|= \partial_{\a\bd}
Z$, where $Z$ is the  complex R-R-scalar of the RNS formulation of the
type II string; its $\t_L\tb_R$-component is $\partial_{\a\ad}
\partial_{\b\bd} S$. The real component of $S$ is the dilaton, its
imaginary component is dual to the antisymmetric tensor of the
NS-NS-sector.
These results can be obtained
by explicit computation from the $\theta$-expansion of the
superfield $V$. After integrating (\ref{Agfin}) and
(\ref{tc}) over chiral and twisted-chiral superspace, respectively, $2g-2$
powers of $F_{\a\b}$ are coupled to two 
 powers of  $C_{\a\b\gamma\delta}$, while $2g-2$ powers of  $\p Z$ are
coupled to two powers of $\partial^2 S$, with the tensorial
structure discussed in \cite{AGNT}. In \cite{BS,BerkovitsJH,RocekIJ} the
question is addressed how 
these (and other) couplings can be described in an off-shell (projective)
superspace description at the non-linearized level.  

\section*{Acknowledgment}
We would like to thank N.~Berkovits, S. Fredenhagen, S.~Kuzenko, 
O.~Lechtenfeld, A. Schwimmer and H.~Verlinde for 
discussions.  This work was partially supported by the RTN contracts
MRTN-CT-2004-503369 and MRTN-CT-2004-005104 and ANR grant BLAN06-3-137168.

\appendix
\section{Conventions and Notations}\label{secConvention}

\subsection{Spinors and superspace}

Throughout this paper we use the conventions of Wess and Bagger \cite{WB}. 
In particular the  
space-time metric is $\eta^{mn}={\rm diag}(-1,+1,+1,+1)$ and the 
spinor metric is $\e^{12}=\e^{\dot 1\dot 2}=\e_{21}=\e_{\dot 2\dot 1}=1$. 
Spinor indices are raises and lowered as $\psi^\a=\e^{\a\b}\psi_\b,\,
\psi_\a=\e_{\a\b}\psi^\b$, and likewise for the dotted indices. 
Spinor indices are contracted in the following way: 
$\psi\chi=\psi^\a\chi_\a,\,\bar\psi\bar\chi=\bar\psi_\ad\bar\chi^\ad$. 
Barred spinors always have dotted indices.  
We define $x_{\a\ad}=\s_{\a\ad}^m x_m$ where 
$\s^m_{\a\ad}=(-{\bf 1},\vec\sigma)$ with $v^m v_m=-\half
v^{\a\ad}v_{\a\ad}$ and $\p_{\a\ad}=\s^m_{\a\ad}\p_m$ such that 
$\p_{\a\ad}x^{\b\bd}=-2\delta_\a^\b\delta_\ad^\bd$.   
Starting from the supersymmetry invariant one-forms on 
superspace
\begin{eqnarray}\label{Es}
e^a&=&d x^a-id\t\s^a\tb+i\t\s^a d\tb\,,\\
\nn e^\a&=&d\t^\a\,,\\
\nn e_\ad&=&d\tb_\ad\,,
\end{eqnarray}
one finds their pullbacks to the world-sheet
\begin{eqnarray}\label{Espb}
\Pi^{\a\ad}&=&\p x^{\a\ad}+2i\p\t^\a \tb^\ad+2i\p\tb^\ad\t^\a\,,\\
\nn \Pi^\a~&=&\p\t^\a\,,\\
\nn \bar\Pi_\ad~&=&\p\tb_\ad\,,
\end{eqnarray}
and likewise for the right-movers. Expressed in terms of the $\Pi$'s,
the energy momentum tensor  
of the $x,\t,p$ variables is
\begin{equation}\label{Txthetap}
T={1\over4}\Pi^{\a\ad}\Pi_{\a\ad}-\Pi^\a d_\a-\bar\Pi_\ad\bar
d^\ad-{1\over2}(\p\rho)^2+{1\over2}\p^2\rho\,,
\end{equation}
with $d_\a$ and $\bar d^\ad$ defined by
\begin{equation}\label{eDefds}
\begin{split}
d_\a&=p_\a+i\tb^{\ad}\p x_{\a\ad}-\tb^2\p\t_\a+{1\over2}\t_\a\p\tb^2\,,\\
 \bar d^{\ad}&=\bar p^\ad+i\t_\a\p
x^{\a\ad}-\t^2\p\tb^\ad+{1\over2}\tb^\ad\p\t^2\,.
\end{split}
\end{equation} 
%
\subsection{Hybrid variables and ${\cal N} =2$ algebra}

The singular parts of the operator products of the hybrid variables are 
\begin{equation}
\label{opehybrid}
\begin{split}
x^m(z,\bar z)x^n(w,\bar w)&\sim-\eta^{mn}\ln|z-w|^2\;,\\
 \t_\a(z)p^\b(w)&\sim{\delta_\a^{\,\,\b}\over(z-w)}\;,\\
 \tb^\ad(z)\bar p_\bd(w)&\sim{\delta^\ad_{\,\,\bd}\over(z-w)}\;,\\
 \rho(z)\rho(w)&\sim-\ln(z-w)\;,
\end{split}
\end{equation}
Both (\ref{Txthetap}) and (\ref{opehybrid}) follow
from the action~(\ref{hybridaction}).
We also note that 
\begin{equation}\label{ddbar}
d_\a(z)\bar d_\ad(w)\sim{2i\Pi_{\a\ad}(w)\over(z-w)}\,,
\end{equation}
while $d d$ and $\bar d\bar d$ are finite. 
The action of $d$ and $\bar d$ on a generic superfield $M$ is 
\begin{equation}
\label{dM}
\begin{split}
d_\a(z) M(w)&\sim-{D_\a M(w)\over(z-w)}\qquad\qquad\hbox{where}\qquad
D_\a\equiv \phantom{+}\p_\a+i\tb^\ad\p_{\a\ad}\\
d_\ad(z) M(w)&\sim-{\bar D_\ad M(w)\over(z-w)}\qquad\qquad\hbox{where}\qquad 
\bar D_\ad\equiv-\pb_\ad-i\t^\a\p_{\a\ad}
\end{split}
\end{equation}
with $\lbrace D_\a,\bar D_\ad\rbrace=-2i\p_{\a\ad}$. For later purposes
we note the useful identities 
\begin{equation}\label{jCovDerIdent}
[D_\a,\bar D^2]=-4i\p_{\a\ad}\bar D^\ad\,\qquad
{1\over16}[\bar D^2,D^2]=\p^m\p_m+{i\over2}\p_{\a\ad}D^\a\bar D^\ad
\end{equation}
One defines the space-time supercharges 
\begin{equation}\label{stSUSYcharges}
  \begin{split}
Q_\a& =\oint \bigl(p_\a-i\tb^\ad\p x_{\a\ad}+{1\over2}\tb^2\p\t_\a\bigr)\\
 \bar Q^\ad&=\oint\bigl(\bar p^\ad-i\t_\a\p
 x^{\a\ad}+{1\over2}\t^2\p\bar\t^\ad\bigr) 
\end{split}
\end{equation}
such as to satisfy 
\begin{equation}\label{stsusy}
  \begin{split}
\lbrace Q_\a,\bar Q_\ad\rbrace&=-2 i\oint\p x_{\a\ad}\\
 \lbrace Q_\a,Q_\b\rbrace&=\lbrace \bar Q_\ad,\bar Q_\bd\rbrace=0
\end{split}
\end{equation}
and to (anti)commute with the $d$'s and $\Pi$'s. In deriving these
relations we have dropped total derivatives involving fermion bilinears.
Note that $\lbrace Q_\a,\bar Q_\ad\rbrace\phi=2i\s^m_{\a\ad}\p_m\phi$, as expected.

Vertex operators for physical states are required to be primary. 
By definition, a primary state is annihilated by all positive modes
of the generators of the superconformal algebra. For a superfield $\cU$ which is
independent of $\rho$ and of the internal CFT this leads to the on-shell
conditions $\p_m\p^m\cU=D^2 \cU=\bar D^2 \cU=0$.

\subsection{The integrated vertex operator}
\label{secVOder}
\noindent We compute  \cite{deBoerKT}
\begin{equation}\label{GGV}
{\cal W}(z)= G^-G^+ \cV(z)
\end{equation}
where $\cV(z)$ is primary (assumed to be bosonic)
and depends on $x,\t,\tb$ but is independent of $\rho$. The first commutator is
straightforward to compute. With the help of (\ref{dM}) one finds
\begin{equation}\label{GV}
\sqrt{32} G^+\cV(w)
=\oint_{C_w} \!\!\!\!\!dz\,e^{-\rho(z)}\,\left(-{\bar D^2
    \cV(w)\over(z-w)^2}
+{2\bar d\bar D\cV(w)\over z-w}\right)=2e^{-\rho(w)}\bar
d_{\dot\alpha}\bar D^{\dot\alpha}\cV(w)
\end{equation}
where $\bar D^2\cV=0$ has been used.
The computation of the second commutator is more involved.
Applying the rules stated in \cite{dFMS} one finds
\begin{multline}\label{GpGmV}
-32 \,G^-G^+\,\cV(w)
 =\\
16\p\tb_\ad\bar D^\ad\Phi(w)+8i d^\a\p_{\a\ad}\bar D^\ad
\cV(w) -4i\Pi_{\a\ad}D^\a\bar D^\ad\cV(w)
+\bar d_\ad D^2\bar D^\ad\cV(w)
\end{multline}
where normal ordering is implied in all terms. A term
$-4i\p\rho\,\p_{\a\ad}D^\a\bar D^\ad\cV(w)$ has been dropped; using
(\ref{jCovDerIdent}) it can be shown to vanish if
$\cV$ is primary, as we have assumed.
One can cast (\ref{GpGmV}) into a more symmetric form if one adds a total
derivative\footnote{This total derivative could contribute to boundary
terms when two vertex operators collide and might thus play an
important role in amplitude computation.} of $\cV$:
\begin{multline}\label{GpGmVsym}
-32 G^- G^+\,\cV(w)+8\p\cV(w)\\
= -8(\Pi^\a D_\a-\bar\Pi_\ad\bar D^\ad)\cV(w)-2i\Pi_{\a\ad}[D^\a,\bar
D^\ad]\cV(w)
+(\bar d_\ad D^2\bar D^\ad-d^\a\bar D^2 D_\a)\cV(w)
\end{multline}
Here we have used the notation defined in (\ref{Espb}).

\section{Mapping the RNS to the hybrid variables}
\label{appMap}

In this appendix we give the details of the field mapping
suppressed in section~\ref{secFR} which relate the RNS and the hybrid
variables, following closely \cite{BerkovitsBF}.  
It is easiest to split this map into a part involving a
field redefinition and one involving a similarity transformation. The
field redefinition defines a set of Green-Schwarz-like variables in
terms of the RNS variables. These are then related
to the hybrid variables by a similarity transformation.

\subsection{Field redefinition from RNS to chiral GS variables}
\label{sec:fr}
From  the RNS variables one first forms a set of variables  according
to (\ref{eMapTheta}) and (\ref{eDefRho}). Following \cite{BerkovitsCY,BerkovitsVY}, these are
called the ``chiral GS-variables".  In
\cite{BerkovitsCY,BerkovitsVY} these variables were denoted
collectively by $\tilde\Phi$, whereas in \cite{BerkovitsBF} they  were
labeled with the superscript ``old". In this section we label the chiral
GS-variables with the subscript ``GS" 
for clarity, while this is suppressed in the main text.   

In order to achieve the correct normalization (\ref{DicSum}) we
must perform the following rescaling of RNS variables:
\begin{equation}\label{jresc}
b\rightarrow 2\sqrt2 b\,,\quad c \rightarrow (2\sqrt{2})^{-1}
c\,,\quad \eta \rightarrow 2\sqrt2\eta\,,\quad \xi\rightarrow
(2\sqrt{2})^{-1}\xi \,,\quad e^{-\phi}\rightarrow 2\sqrt2e^{-\phi}\,.
\end{equation}
These rescalings preserve all the OPEs. We use the rescaled RNS
variables in this section.

\subsection{Similarity transformation relating chiral GS to hybrid variables}
\label{sec:simtr}

The chiral GS-variables $\Phi_{\rm GS}$, including those of the 
internal SCFT, are related to the hybrid ones $\Phi$  by the
similarity transformation
\begin{equation}\label{jFieldRedef}
(\Phi)_{{\rm GS}} = e^{{\cal M}+{\cal M_C}^-}( \Phi)\, 
e^{-({\cal M}+{\cal M_C}^-)}\,.
\end{equation}
We have defined 
\begin{equation}\label{ja}
{\cal M} =\oint i \t^\a\tb^\ad\p x_{\a\ad} +{1\over 4}
(\t^2\p\tb^2-\tb^2\p\t^2)\,, 
\end{equation}
and 
\begin{equation}\label{jb}
{\cal M}_C^- = - \sqrt2 c_-\oint e^{-\rho} \t^2  G_C^-\,,\quad
{\cal M}_C^+= \sqrt2 c_+ \oint e^{\rho} \tb^2 G_C^+\,,
\end{equation}
with ${\left[{\cal M}_C^\pm , {\cal M}\right] = 0}$. One way to see
that this is indeed the correct transformation is to verify that  
\begin{equation}\label{jc}
e^{\cal M} p_\a e^{-\cal M} =  d_{\a} \,,
\end{equation}
from which
\begin{equation}\label{jppb}
e^{{\cal M} + {\cal M}_C^-}\left({1\over\sqrt{32}} e^{\rho}p^\a
p_\a\right) e^{-({\cal M} + {\cal M}_C^-)}
  = {1\over\sqrt{32}}e^{\rho} d^\a d_\a + c_- G_C^-={\cal G}^- 
  \end{equation}
follows. By definition~(\ref{jFieldRedef}), the l.h.s. of this expression equals
$({1\over\sqrt{32}}e^{-\rho} p^\a p_\a)_{{\rm GS}}$, which, according
to (\ref{eMapTheta}) and (\ref{jresc}), equals the RNS ghost field
$b$. One therefore concludes
\begin{equation}
\label{eq:wasweissich}
  b = ({1\over\sqrt{32}}e^{-\rho} p^\a p_\a)_{{\rm GS}} =
  {1\over\sqrt{32}}e^{-\rho} d^\a d_\a+ c_-G_C^- = \cG^-
\end{equation}
as stated in
(\ref{eThTRNS}). 

It can be verified that for the generators  $
\cJ = \cJ_{{\rm GS}}$,  $\cT =\cT_{{\rm GS}}$, ${\cal J}^{\pm\pm}
=\cJ^{\pm\pm}_{\rm GS}$. They are therefore not affected by
(\ref{jFieldRedef}). These results were used in
section~\ref{secNfour}. 

\subsection{Hermitian conjugation of the hybrid variables}
Hermitian conjugation acts on the hybrid variables as
\begin{equation}\label{herm}
({x^m})^\dagger = x^m\,, \quad (\t^\a)^\dagger = \tb^\ad\,,\quad 
(p_\a)^\dagger = -\bar p_\ad\,.
\end{equation}
From these properties one concludes that $(\p x^m)^\dagger = - \p x^m$ and 
$(\p \t^\a)^\dagger = -\p\tb^\ad$. In addition, we define
\begin{equation}
\rho^\dagger=-\rho-\ln z+i\pi\,,\quad H^\dagger=H-i\sqrt{3}\ln
z+\pi\sqrt{3}\,.
\end{equation}
Some comments are in order here.  The $\ln z$ terms
are due to the background charges of currents $J=\p\rho$ and
$J_C=i\sqrt{3}\p H$. In the presence of a (real) background charge
$Q$, the operator product of the energy-momentum tensor and a generic
(hermitian) current is modified to
$T(z)j(w)\sim{Q\over(z-w)^3}+{j(w)\over(z-w)^2}+{\p j(w)\over(z-w)}$.
In terms of the modes this reads
$[L_n,j_m]={1\over2}Qn(n+1)\delta_{n+m}-m j_{n+m}$ and implies
$j_n^\dagger=j_{-n}-Q\delta_{n,0}$ and $L_n^\dagger=L_{-n}-Q(n-1)
j_{-n}$.  These results are to be applied for the currents $j= -\p\rho = -J$ and
$j=J_C$ with the background charges $-1$ and $-3$, respectively. This
implies that $(p_\rho)^\dagger = p_\rho-1$ and $(p_H)^\dagger =p_H
+\sqrt3$ such that the cocycle factors introduced in section
\ref{sec:hyb} satisfy $(c_+)^\dagger=c_-$. The
constant shifts $+ i\pi$ and $+\pi\sqrt3$ seem to be needed in order to obtain the 
correct hermiticity relations between various ${\cal N}=4$ generators
and the correct algebra. It is consistent with the fact that $\rho$
and $H$ are compact bosons with periodicity $2\pi i$ and
$2\pi\sqrt{3}$, respectively.  Using the general CFT rule,
$[\phi(z)]^\dagger=\phi^\dagger({1\over\bar z})\bar z^{-2 h}$, valid
for a primary field $\phi$ of dimension $h$, one shows
$\exp(q\rho)^\dagger=(-1)^q \exp(-q\rho)$ and
$\exp({iq\over\sqrt{3}H})^\dagger=(-1)^q \exp({-{iq\over\sqrt{3}}H})$.
This, together with $(G_C^\pm)^\dagger = G_C^{\mp}$, completes the
discussion of hermitian conjugation of the hybrid variables. 

\subsection{Hermitian conjugation of the RNS variables}

Going through the sequence of similarity transformations
and field redefinitions outlined in Appendices~\ref{sec:fr} and
\ref{sec:simtr}, the hybrid conjugation rules induce a hermitian
conjugation for the RNS variables. This conjugation is not the
standard one. We discuss this in detail below and obtain,  as a
side-product, the a justification for the complete 
dictionary given in~(\ref{DicSum}).

Using ${\cal
M}^\dagger = {\cal M}$ and
$({\cal M}_C^-)^\dagger = 
{\cal M}_C^+$, hermitian conjugation of~(\ref{jppb}) and~(\ref{eq:wasweissich})
(or direct computation) yields  
\begin{equation}\label{jjbrstG}
b^{\dagger} = e^{-({\cal M} + {\cal M}_C^+)}\left(- {1\over \sqrt{32}}
e^{-\rho}\bar p_\ad \bar 
p^\ad\right) e^{{\cal M} + {\cal M}_C^+} 
=- {1\over \sqrt{32}} e^{-\rho} \db_\ad \db^\ad -c_+  G_C^+= 
{\cal G}^+\,.
\end{equation}
As is argued below, this expression equals the current $j_{\rm BRST}$
in accordance with (\ref{eThTRNS}). The hybrid hermiticity properties
therefore imply in particular that $b^\dagger = j_{\rm BRST}$. We work
this out in more detail: one first remarks that~(\ref{jjbrstG}) is not
the similarity transformation~(\ref{jFieldRedef}), since latter
involves the charge ${\cal M} + {\cal M}_C^-$. In fact, under this
transformation $-{1\over\sqrt{32}}e^{-\rho}\bar p^2$ is mapped to
\begin{equation}
e^{{\cal M} + {\cal M}_C^-}
\left(-{1\over\sqrt{32}}e^{-\rho}\bar p_\ad \bar 
p^\ad\right) e^{-({\cal M} + {\cal M}_C^- )}=  
\left( - {1\over\sqrt{32}} e^{-\rho}\bar p_\ad \bar p^\ad\right)_{\rm
GS} = -b \gamma^2\,.
\end{equation}
The first equality is just the definition~(\ref{jFieldRedef}), while
the second one is a consequence of the field
redefinition~(\ref{eMapTheta}) and (\ref{jresc}). Inverting this 
relation and inserting the result in (\ref{jjbrstG}) one finds
\begin{equation}\label{jbrstfrombgg}
b^\dagger = e^{-\cal R} (-b\gamma^2 )e^{{\cal R}}.
\end{equation}
The claim is that the r.h.s. is $j_{\rm BRST}$. We have
defined 
\begin{equation}\label{jR}
e^{\cal R}=e^{{\cal M}+{\cal M}_C^-} e^{{\cal M}+{\cal M}_C^+}
=e^{2{\cal M}+{\cal M}_C^-+  
{\cal M}_C^++{1\over 2}[{\cal M}_C^-,{\cal M}_C^+]}\,.
\end{equation}
While the first equality is the definition, the second one holds
whenever one has  
$[{\cal M}_C^-,[{\cal M}_C^-,{\cal M}_C^+]]=[{\cal M}_C^+,[{\cal
M}_C^+,{\cal M}_C^-]] = 
0$. That this is indeed the case as can be seen by calculating the commutator
\begin{equation}\label{jco}
[{\cal M}_C^-,{\cal M}_C^+]  =
2 \oint \left[ \t^2\tb^2 \left(J_C + {c\over3}\p\rho\right)-
{c\over 3}\tb^2 \p\t^2 \right]\,.
\end{equation}
We used the normalization
\begin{equation}\label{jGG}
G_C^-(z) G_C^+(w) \sim {{c\over 3}\over(z-w)^3} -
{J_C(w)\over(z-w)^2} 
+{T_C(w)-\p J_C(w)\over z-w}\,,
\end{equation}
which follows from (\ref{GplusGminus}).
Since $G_C^\pm$ has only a simple pole with $J_C$,  ${\cal
M}_C^\pm$ commute with this commutator and (\ref{jR}) is established. The
explicit expression for ${\cal R}$ in terms of hybrid variables is  
\begin{equation}
\label{eq:R}
\begin{split}
{\cal R}  &= \oint \left[2 i \t^\a \tb^\ad \p x_{\a\ad} - \sqrt2 c_-e^{-\rho}\t^2
G_C^-+\sqrt2 c_+e^{\rho}\tb^2  G_C^+ \right.\\ &\left.
\qquad\qquad
 +\t^2\tb^2(J_C+{c\over 3} \p\rho) +
(1+{c\over 3})\t^2\p\tb^2\right]\,.
\end{split}
\end{equation}
In order to evaluate the r.h.s. of~(\ref{jbrstfrombgg}), we
re-express this operator in terms of RNS variables. Thereby one must bear in 
mind that the field map (\ref{jFieldRedef}) affects all
fields, including the generators of the internal SCFT. In particular, one
finds that under (\ref{jFieldRedef}) 
\begin{equation}
\begin{split}
(\p x_{\a\ad})_{\rm GS} &=  \p x_{\a\ad}+ 2i \p(\t_\a\tb_\ad)\,,\\
(c_+e^{\rho} \tb^2 G_C^+)_{\rm GS} & = c_+e^{\rho} \tb^2 G_C^+ +
\sqrt2\left[\t^2\tb^2\left(J_C+{c\over 3}\p\rho
\right)-{c\over3}\tb^2\p\t^2\right]\,, 
\end{split}
\end{equation}
while $\t^\a$, $\tb_\ad$, and the combinations $\t^2\tb^2 J_C$ and
$\t^2\tb^2\p\rho$ remain unaffected. Using~(\ref{eNewJC}),
(\ref{eNewGenerators}), and (\ref{eDefRho})
and dropping total derivatives one finds that ${\cal R}$ is the following simple
expression in RNS variables:\footnote{In order to obtain this result,
  one must take special care of the overall
   signs for the RNS expressions of $\t^2$, $\tb^2$,  $\t^2\p\tb^2$, and
   alike. We suppress these details in this note.}
\begin{equation}\label{jRRNS}
{\cal R} =\oint \left[c\xi e^{-\phi} T_F+{1\over2}e^{-2\phi} c\p
c \xi\p\xi(\p\phi+{c-9\over 3} \p\sigma)\right] \,.
\end{equation}
The terms in (\ref{eq:R}) involving the current $\breve J_C$ have
canceled and the last term in (\ref{jRRNS}) vanishes for $c=9$. We have
defined 
$T_F = T_F^{x,\psi} + \breve G_C^+ + \breve G_C^-$,
where $T_F^{x,\psi}$ is the supercurrent of the space-time matter
sector. It is normalized as $T_F(z) T_F(w) \sim 
{2\over3}(c^{x,\psi}+c)(z-w)^{-3} +\ldots$, with $c^{x,\psi} = 6$ [see also
(\ref{jGG})]. 

Using the conventions of~\cite{FMS} one can verify that
(\ref{jbrstfrombgg}) with ${\cal R}$ given in (\ref{jRRNS}) indeed
produces the BRST current (this current differs from the usual current
by addition of total derivative terms),
\begin{equation}
\begin{split}
 e^{-\cal R}(-b\gamma^2 )e^{\cal R}=  j_{\rm BRST} 
=&\, c\left( T-b\p c-{1\over
2}(\p\phi)^2-\p^2\phi+{1\over2}(\p\chi)^2+{1\over2}\p^2\chi\right)\\
\label{jbrstres} & + \gamma\, T_F-b\,\gamma^2+\p^2c+\p(c\p\chi)\,. 
\end{split}
\end{equation}
where $T = T^{x,\psi} + \breve T_C$. Details of this computation can be found
in \cite{AcostaHI}. The BRST charge is $Q_{\rm BRST} = \oint \left[c
T+ e^\phi\eta \,T_F  +b c\p c+ b e^{2\phi}\eta\p\eta  +
c(\p\xi\eta-{1\over2}(\p\phi)^2-\p^2\phi)\right]$ and coincides with
the charge which follows from the BRST current
in \cite{PolchinskiRR} after bosonization. From this one derives
an expression for the picture-changing operator in 
bosonized form,
\begin{equation}\label{picchop}
Z=\{Q_{\rm BRST},\xi\} = e^\phi T_F+c\p\xi - b\p\eta
e^{2\phi} - \p(b\eta^{2\phi})\,,
\end{equation}
which enters in~(\ref{eDic}).

It does not seem possible to give a closed formula for the hermitian
conjugation of a generic RNS field. If, however, an RNS field
$\Phi_{\rm RNS}$ is expressible in terms of chiral GS-variables,
$\Phi_{\rm GS}=\Phi_{\rm RNS}$, one can use the same argument as above
and deduce the rule:
\begin{equation}\label{jhermrns}(\Phi_{\rm RNS})^\dagger = e^{-\cal
    R}\, \Psi_{\rm RNS}
\, e^{\cal R} \,,\quad\hbox{with}\quad \Psi_{\rm RNS} :=
(\Phi_{\rm GS})^{\dagger}\,,
\end{equation}
where $(\Phi_{\rm GS})^\dagger$ is calculated the same way as the
corresponding expression in hybrid variables. For instance, in above argument,
$\Phi_{\rm RNS} = b = \left({1\over\sqrt{32}} e^{-\rho} p^2\right)_{\rm GS} =
\Phi_{\rm GS}$, and $\Psi_{\rm RNS} = (\Phi_{\rm GS})^\dagger =
\left(-{1\over\sqrt{32}} e^{\rho} \bar{p}^2\right)_{\rm GS} = - b \gamma^2
$, which leads to~(\ref{jbrstfrombgg}).

Some clarifying remarks on the hermitian conjugation rule of RNS
variables are in place here. The conformal weights of $\Phi_{\rm RNS}$
and $\Psi_{\rm RNS}$ generally differ when evaluated w.r.t.  $T_{\rm RNS}$. 
The reason for this is the following: 
as we have reviewed above, in the presence of a background charge, 
$T^\dagger=T-Q\p j$. If ${\cal O}$ is an operator with ${\rm U}(1)$ charge
$q$, it can be written in the form 
${\cal O}=\exp(q\int^z\!\!j){\cal O}'$ with ${\cal O}'$ neutral under $j$.
(Here we have normalized the current according to $j(z) j(w)\sim
{1\over (z-w)^2}$). The hermitian conjugate operator is 
${\cal  O}^\dagger=\exp(-q\int^z\!\!j)({\cal O}')^\dagger$.  Its
conformal weight measured with $T^\dagger$ is the same as that of
${\cal O}$ measured with $T$. One defines the operator
$\widetilde{{\cal O}}=\exp(q\int^z\!\!j)({\cal O}')^\dagger$ which
has the same ${\rm U}(1)$ charge and weight (w.r.t. $T$) as ${\cal O}$.

For the case of interest, this means that 
$T=-{1\over2}(\p\rho)^2-{1\over2}(\p 
H)^2+{1\over2}\p^2\rho+{i\over2}\sqrt{3}\p^2H$   
becomes $T^\dagger=T-\p^2\rho-i\sqrt{3}\p^2H=T-\p{\cal J}$. 
The conformal weight of $\Psi_{\rm RNS}$ w.r.t. $T^\dagger$ is then the
same as that of  
$\Phi_{\rm RNS}$ w.r.t. $T$. It is now straightforward to find
that the $\Psi_{\rm RNS}$ corresponding to $\Phi_{GS}= e^\sigma$,
$e^\chi$, and $ e^\phi$, for example are given (up to overall signs and
rescalings) by $e^{\sigma +2\chi-2\phi}$, $e^{2\sigma+\chi-2\phi}$, and
  $e^{2\sigma+2\chi-3\phi}$, respectively.
Finally, we define the operator conjugation ${\cal O}\rightarrow
\widetilde{{\cal O}}$ for the case at hand. We write any operator with $\rho$-charge $p$ (with respect to the current $\p\rho$)
and ${\rm U}(1)_C$-charge $q$ as
\begin{equation}\label{eOO}
{\cal O}=e^{-p\rho+{iq\over\sqrt{3}}H}\,{\cal O}'
=e^{{1\over2}(p+q)\int^z{\cal
J}}\, e^{-{1\over2}(3p+q)(\rho+{i\over\sqrt{3}}H)}\,{\cal
O}'\,.
\end{equation}
Then an operator conjugation preserving the conformal w.r.t. to $T$ is
defined by  
\begin{equation}\label{eOs}
\widetilde{\cal O}=e^{{1\over2}(p+q)\int^z{\cal J}}\,
e^{{1\over2}(3p+q)(\rho+{i\over\sqrt{3}}H)}\,({\cal O}')^\dagger
= e^{(2p+q)\rho+ (3p+2q)\tfrac{i}{\sqrt3}H}(\cal O')^\dagger\,. 
\end{equation}
This is the conjugation used in section~\ref{sec:hyb}.

\section{Vertex operators} 
\label{secMap}

\subsection{Massless RNS vertex operators}

The field redefinition between the RNS and hybrid variables presented in
section~\ref{secFR} induces a map of the vertex operators of the RNS formulation
to those of the hybrid formulation. We first discuss the unintegrated vertex
operators for massless states. The field redefinition~(\ref{eMapTheta}) [or
(\ref{eRTheta}) for the right-moving sector of the type IIA string] relates the RNS
vertex operators in the large Hilbert space to operators expressed in terms of
chiral GS-variables. To obtain the vertex operators in the hybrid
variables one needs to perform the additional map~(\ref{jFieldRedef}). It can be
shown, however, that this map does not affect any of the expressions discussed
below. The reason is that at the massless level the unintegrated vertex
operators do not depend on $p$ or $\bar p$. Furthermore, they contain at least
two powers of $\t$ and $\tb$ such that the map is trivial as long as the
internal part of the vertex operators are primary.

The vertex operators of the bosonic
components of the
space-time ${\cal N}=2$ multiplets descend from the NSNS and RR sectors 
of the 10d superstring. The vertex operators in the
large Hilbert space are of the general form 
\begin{equation}\label{eVOPdef}
\cV^{(q,\tilde q)}= |c \xi e^{\a\phi}|^2\,
\breve\Phi^{(q,\tilde q)}\, W\,. 
\end{equation}
As explained in subsection~\ref{secNfour}, $\cV$ has conformal weight
0 and ghost number 0 with respect to~(\ref{eRNSghostC}). $W$ is the
space-time part of the vertex operator; $\Phi^{(q,\tilde q)}$ are
primary fields of ${\rm U}(1)_L\times {\rm U}(1)_R$ charge $(q,\tilde
q)$ of the internal $c=9$, ${\cal N}=(2,2)$ SCFT. In what follows, we
mainly concentrate on the left-moving part of the vertex operators,
which we denote by $\cV^{(q)}$.  Notice that the charge of
$\breve\Phi^{(q)}$ with respect to $\breve J_{C}$ is the same as the
one of $\Phi^{(q)}$ defined by~(\ref{ePicTw}) with respect to $J_{C}$
of~(\ref{eNewJC}). According to (\ref{eDefRho}), the charge of the
vertex operators $\cV^{(q)}$ under $J=\p\rho$ is $-q+ 3\, (1+\alpha)$
and must therefore carry a factor of
$e^{\left[q-3(1+\alpha)\right]\rho}$ when expressed in hybrid
variables. Since the ghost number $J_{\rm gh} = \cJ = \p\rho+J_C$ of
the vertex operator $\cV^{(q)}$ is zero, the $\rho$-charge is minus
the $J_C$-charge.  Therefore the internal part of $\cV^{(q)}$ in
hybrid variables must involve $\Phi^{(q-3[1+\a])}$. Given any RNS
vertex operator, this rule fixes the form of the vertex operator in
the hybrid formulation up to the spacetime part. The latter is
determined by (\ref{eMapTheta}) from which one derives, for
example,\footnote{Here and in what follows we suppress overall signs
  and numerical factors.}
\begin{equation}\label{eERho}
e^{-\rho} \, \theta^2=c\,, \qquad 
e^{\rho}\, \bar\theta^2= c \, e^{-2(\phi-\chi)}\,.
\end{equation}
The RNS vertex operators are restricted by the requirement that their 
operator product with the spacetime gravitino $e^{-\half\phi}S^\a\Sigma$
is local. This applies to the left-moving part. For the right-moving part of 
type IIB (IIA) locality with $(e^{-\half\phi}S^\a{\Sigma})_R$ 
($(e^{-\half\phi}{S}^\a{\bar{\Sigma}})_R$) is required. 
Given the OPE $\Sigma(z)\breve\Phi^{(q)}(w)\sim
(z-w)^{q\over2}\breve\Phi^{(q+{3\over2})}(w)+\dots$ 
this implies restrictions on $q$. 

It is now straightforward to find the following maps of 
vertex operators in the NS sector in the canonical ghost picture 
($\a=-1$):\footnote{We do not display the
$e^{i\, k\cdot X}$ factors  which must be included in the complete
expression for each vertex.}
\begin{equation}\label{eMapNS}
  \begin{split}
\cV_{\rm NS}^{(0)} &= c\,\xi \,e^{-\phi}\,\psi^m =
(\theta\sigma^{m}\bar\theta) \,,\\
 \cV_{\rm NS}^{(+1)}& = c\,\xi \, e^{-\phi}\, \breve\Phi^{(+1)} = e^{\rho}\,
\bar\theta^2\, \Phi^{(+1)}\,,\\
 \cV_{\rm NS}^{(-1)}&= c\,\xi \, e^{-\phi}\, \breve\Phi^{(-1)} = e^{-\rho}\,
\theta^2\, \Phi^{(-1)}\,.
\end{split}
\end{equation}
In the R sector in the canonical ghost 
picture ($\alpha=-{1\over2}$) one finds, for example, 
\begin{equation}\label{eMapR}
\begin{split}
\cV_{\rm R}^{(+{3\over2})}&=c\,\xi \, e^{-{\phi\over2}}\, S^{\alpha}\Sigma
= \theta^{\alpha}\, \bar\theta^2\,,\\
\cV_{\rm R}^{(+{1\over2})}&=c\,\xi \, e^{-{\phi\over2}}\,  \bar
S^{\dot\alpha} \, \breve\Phi^{(+{1\over2})}  
= e^{-\rho}\,\bar\theta^{\dot\alpha}\,\theta^2\, \Phi^{(-1)}\,,\\
 \cV_{\rm R}^{(-{1\over2})}&=c\,\xi \, e^{-{\phi\over2}}\, S^{\alpha}\,
\breve\Phi^{(-{1\over2})} 
= e^{-2\rho}\, \theta^{\alpha}\,\p \theta^2\, \Phi^{(-2)}\,.
\end{split}
\end{equation}
These expressions illustrate that RNS vertex operators, which are
(RNS) hermitian conjugates of each other, are generally not mapped to
operators which are hermitian conjugates in the hybrid sense
(cf.~Appendix~\ref{appMap}). Conversely, two hermitian conjugate
hybrid operators are related to RNS operators in different ghost
pictures. For instance, $\bar\t^\ad \t^2(w)=c\p c\xi\p\xi
e^{-{5\over2}\phi}\bar S^\ad\bar\Sigma(w)= \lim_{z\to w}Y(z)\, c\xi
e^{-{\phi\over2}}\bar S^\ad\bar\Sigma(w)$ where $Y=c\p\xi e^{-2\phi}$
is the inverse picture changing operator.

For type IIB compactifications (\ref{eERho}), (\ref{eMapNS}) and (\ref{eMapR})  are the same
for both the left- and right-moving sectors. For type IIA
compactifications the expressions for the right-movers are different;
they can be obtained from above relations by reversing the signs of all
explicit charge labels and replacing $\Sigma \leftrightarrow
\bar{\Sigma}$.   
%

\subsection{Universal massless multiplets}

The vertex operators for the universal sector of type II strings on
CY${}_3$ \cite{rfCFG,AntoniadisSW} are associated to the identity
$\Phi^{(0,0)}=\uno$ and the states in the RR sector connected to the
identity by spectral flow. They are grouped in to the real superfield
${\cal U}(x,\t_{L,R},\tb_{L,R})$, which was constructed in
\cite{BerkovitsCB}, and contains the $24+24$ degrees of freedom of
supergravity multiplet and the $8+8$ degrees of freedom of the
universal tensor multiplet (which can be dualized to the universal
dilaton multiplet). In the Wess-Zumino gauge, the metric, the
antisymmetric tensor, and the dilaton appear at the lowest non-vanishing
order of the $\t$-expansion of ${\cal U}$. Other fields, such as
the (anti)selfdual part of the graviphoton field strength $F_{\a\b}$
($F_{\ad\bd}$) and the derivative of the complex RR-scalar $Z$, to
which we referred to in the main text, appear at higher orders in the
$\t$-expansion:
\begin{equation}\label{Uuniv}
  \begin{split}
{\cal U} =& \,\zeta_{mn}(\t\s^m\tb)_L(\t\s^n\tb)_R + \left[ F_{\a\b} \t_L^{\vphantom{\beta}\a}\t_R^\b |\tb^2|^2 + {\rm h.c.}\right]\\
   & + \left[(\p_{\a\bd} Z +\ldots ) \t_L^{\vphantom{\alpha}\a}
     \tb^2_L \tb_R^\bd \t_R^2 + {\rm h.c.}\right] +\ldots
  \end{split}
\end{equation}
The full expansion can be found in \cite{BerkovitsCB}.
Using the expressions~(\ref{eMapNS}) and~(\ref{eMapR}) these operators are
identified as the RNS vertex operators. One finds, for instance, 
\begin{equation}\label{eVOPuniv}
  \begin{split}
{\cal U}_{\rm \zeta} &=\zeta_{mn}\, \psi^{m}_L {\psi}^n_R\,
\left| c\,\xi e^{-\phi} \right|^2\,\\
{\cal U}_{\p Z} &=\p_{m} Z\,(S_L \sigma^{m}\bar
S_R)\, \left| c\,\xi \, e^{-\half\phi} \Sigma\right|^2\\ 
 {\cal U}_{\p\bar{Z}}' &=\p_m\bar Z\, (\bar S_L
\sigma^{m}S_R) \left| c\,\xi\,e^{-\half\phi} \bar\Sigma\right|^2
      \end{split}
\end{equation} 
As explained in the previous section, it is not ${\cal U}_{\p \bar Z}'$ that is
mapped directly to the hybrid vertex operator, but the picture changed
operator $Y{\cal U}_{\p\bar Z}'$. 

\subsection{Compactification dependent massless multiplets}
\label{sec:chirtwichir}
The spacetime parts of the massless vertex operators, the presence of
which depends on the
particular choice of Calabi-Yau compactification, can be grouped into
real chiral or twisted-chiral multiplets as described in sec.~\ref{sec:maveop}.
 
Chiral superfields $M_c$ satisfy $\bar D_{\ad L} M_c = 0 = \bar D_{\ad
  R} M_c$.  Real chiral superfields (vector multiplets) satisfy in
addition $D^2_L M_c = \bar D^2_R \bar M_c$ and comprise $8+8$
components. The chirality constraint means that $M_c$ is a function of
$y^m=x^m+i (\t\sigma^m\tb)_L+i (\t \sigma^m\tb)_R$,
$\t_L$ and $\t_R$ where $y^{m}$ satisfies $D_L y^{m}=\bar{D}_R y^{m}=0$. Parts of the $\t$-expansion are 
\begin{equation}
M_c(y^m,\t_L,\t_R)=t+\cdots+ f_{\a\b}\t^{\vphantom{\b}\a}_L\t^\b_R+\cdots+|\t^2|^2\p^m\p_m\bar t'\,,
\end{equation}
where the complex scalars $t$ and $t'$ and the selfdual two-tensor
$f_{\a\b}$ are functions of $y^m$. The reality constraint implies in particular
that $t=t'$ and that the two-tensor satisfies the Bianchi constraints, which are
solved by writing it as a vector field strength. The complete
expansion can be found in \cite{BerkovitsCB}.  

Twisted-chiral superfields $M_{tc}$ satisfy $\bar D_{\ad L} M_{tc} = 0
= D_{\a R} M_{tc}$. Real twisted-chiral superfields (tensor
multiplets) satisfy in
addition $D^2_L M_{tc} =D^2_R \bar M_{tc}$ and comprise $8+8$
components. The relevant parts of its expansion are 
\begin{equation}
  M_{tc}(z, \t_L,\tb_R)= l_{++}+\cdots
  +v_{\a\bd}\,\t_L^{\vphantom{\bd}\a}\tb_R^{\bd}+\cdots+
  \theta^2_L\,\tb_R^2\,\partial^m\partial_m l_{--}\,. 
\end{equation} 
where $v_{\a\bd}=v_m\sigma^{m}_{\a\bd}$ is a complex vector and
$l_{\pm\pm}$ complex scalars.  All component fields are functions of
$z^m=x^m+i(\t\sigma^m\bar\t)_L-i(\t\sigma^m\bar{\t})_R$ with $\bar D_L
z^m=D_R z^m=0$.  The reality condition implies $\bar l_{++}=l_{--}$.
Its real part requires $\p_m v_n-\p_n v_m=0$ while for its imaginary
part we need $\p^m v_m=0$. These conditions are solved for $v_m=\p_m
l_{+-}+i\epsilon_{mnpq}H^{npq}$ with $H=dB$.  The three scalars
($l_{+-}, l_{--}=\bar l_{++}$) form a ${\rm SU}(2)$ triplet.  The
antisymmetric tensor with field strength $H$ can be dualized to
a fourth scalar which can be combined with $l_{+-}$ to a complex
scalar.  The complete expansion of this field can again be found in
\cite{BerkovitsCB}.


\subsubsection{K\"ahler moduli}
\label{sKaehler}

The $h^{1,1}$ complexified K\"ahler deformations are in one-to-one
correspondence to elements of $H^{1,1}(CY_3)$. In the CFT description
they are described by twisted-chiral primaries
$\Omega_{tc}$ in the $(c,a)$ ring of charge $q_L=-q_R= 1$ and
conformal weight $h_L=h_R={1\over2}$ (in the untwisted theory).
They are obtained, via 
spectral flow, from RR ground states with $q_L=-q_R=-{1\over2}$ and
$h_L=h_R={3\over8}$.  In type IIA these deformations are associated
with the complex scalars of vector multiplets and for type IIB with
the NSNS-scalars of hypermultiplets (or tensor multiplets). 

Here, we focus on type IIA, but the generalization to type IIB is
straightforward.  The corresponding hybrid vertex operators were
given in~(\ref{eVOIIA}).
\begin{equation}
\label{eVOPKaehler}
{\cal U}_{ca} = |e^\rho\tb|^2 M_c\Omega_{tc}\,.
\end{equation}
Note that for the twisted type IIA theory $\Omega_{tc}$ has conformal
weight $h_L= h_R = 0$ (while ${\bar\Omega}_{tc}$ has conformal
weight $h_L= h_R = 1$) such that $M_c$ indeed describes massless states. 
In the large volume limit the
twisted-chiral primary operators can be written 
as\footnote{The left-moving $(\psi_L^i,\psi_L^{\bar\imath})$ are
  twisted to $(\chi_L^i,\lambda_L^{\bar\imath})$ with conformal weights
  $(0,1)$. For the type IIA twist the right-movers
  $(\psi_R^i,\psi_R^{\bar\imath})$ are twisted to
  $(\lambda_R^i,\chi_R^{\bar\imath})$ with conformal weight
  $(1,0)$.}
\begin{equation}\label{eOmegatc}
\Omega_{tc}=h_{i\bar\jmath}\, \chi_L^{i}
{{\chi}_R^{\bar\jmath}}\,,\quad {\bar\Omega}_{tc} =
h_{i\bar\jmath} \lambda_L^{\bar\jmath}\lambda_R^{i}\,.
\end{equation}
Here $h_{i\bar\jmath}$ is an element  of $H^{1,1}(CY_3)$. For notational 
simplicity we drop the additional index which distinguishes between
the $h^{1,1}$ different elements. 

The bosonic degrees of freedom of the
$h^{1,1}$ vector multiplets are found by expanding $M_c$ in powers
of $\t_L$ and $\t_R$,
\begin{eqnarray}\label{eVOPint}
{\cal U}_{t}&=& t\,
\, |e^{\rho}\tb^2|^2\,  {\Omega}_{tc}\,,\nonumber\\
{\cal U}_{f} &=& f_{\a\b}\,\t^{\vphantom{\b}\a}_L\t^\b_R
\, |e^{\rho}\tb^2|^2\,{\Omega}_{tc}\,.
\end{eqnarray}
For each K\"ahler modulus there is a complex
polarization $t$. Its 
real (imaginary) part is the  
space-time scalar field corresponding to fluctuations of the internal 
NSNS $B$-field $B_{i\bar\jmath}$ (mixed components of the CY metric
$g_{i\bar\jmath}$). $f_{\a\b}$ is the selfdual part of a field
strength of the vector 
multiplet's gauge field which arises from  the  
reduction of the three-form potential. This can be seen by relating
these expressions to the vertex operators in the RNS
formulation. Using
(\ref{eMapNS}) and (\ref{eMapR}) one 
finds 
\begin{eqnarray}\label{eVOPKaehlRNS}
{\cal U}_t &=& t\, | c\,\xi e^{-\phi}|^2 \;\breve\Omega_{tc}\,,\\ 
\nn ({\cal U}_t)^\dagger &= &\bar t\, |c\,\xi
e^{-\phi}|^2\;\bar{\breve\Omega}_{tc}\,,\\ 
\nn ({\cal U}_f)^\dagger &=& f_{\ad\bd}
\bar S^{\ad}_L\, {\bar S}^{\bd}_R | c\,\xi e^{-\half{\phi}}|^2\;\bar{\breve{\Omega}}_{tc} \,.
\end{eqnarray}
Incidentally, the choice in (\ref{eMapTheta}) is such that $({\cal
U}_{f})^\dagger$ is mapped to a simple RNS vertex 
operator in the canonical ghost picture while
${\cal U}_{f}$ is mapped to a RNS operator in another ghost
picture.

\subsubsection{Complex structure moduli}

The $h^{2,1}$ complex structure deformations
are related to chiral primary fields $\Omega_c$ in the chiral
$(c,c)$ ring of charge $q_L= q_R= 1$ and conformal weight $h_L=
h_R={1\over2}$ (in the untwisted theory). These are related to operators
describing RR ground states with charges $q_L=q_R=\pm{1\over2}$ and
$h_L=h_R={3\over8}$ by spectral flow.  Again, we focus on the type
IIA string, in which case they correspond
to the NSNS scalars in hypermultiplets, other than the universal one
which contains the dilaton. In the hybrid formalism the space-time
part of these states is described by a real twisted-chiral multiplet
$M_{tc}$ with the field content of a tensor multiplet. The vertex
operators are contained in the potentials given in~(\ref{eVOIIA})
\begin{equation}\label{eVOPOneTwo}
{\cal U}_{cc}= e^{\rho_L-\rho_R}\, \tb^2_L\, {\t}_R^2\,
M_{tc} {\Omega}_{c} \,.
\end{equation} 
The chiral primary field $\Omega_c$ has conformal weight $h_L=0$ and
$h_R=1$ (while ${\bar\Omega}_c$ has weights $h_L=1$ and $h_R=0$) such
that $ M_{tc}$ indeed describes massless states. In the 
large volume limit one has
\begin{equation}\label{eOmegac}
\Omega_c = h_{ij}\chi_L^{i}\lambda_R^{j}\,,\quad  
{\bar\Omega}_c =
h_{\bar\imath\bar\jmath}\lambda_L^{\bar\imath}\chi_R^{\bar\jmath}\,,
\end{equation}
where $h_{ij}= g_{j\bar\jmath} h_{i}{}^{\bar\jmath}$, and
$h_{i}{}^{\bar\jmath}$ is related to 
elements $Y_{i\bar\jmath\bar k}=
h_{i}{}^{\bar\imath}\bar{\Omega}_{\bar\imath\bar\jmath\bar k}$ of
$H^{1,2}(CY_3)$. Again, we suppress the index that distinguishes between
these $h^{2,1}$ different elements. 

The vertex operators contained in this multiplet can be extracted by expanding 
$M_{tc}$ in powers of $\t_L$ and $\tb_R$. The lowest components are
\begin{equation}\label{eVOPhh}
\begin{split}
{\cal U}_{l_{++}}& =l_{++}
e^{\rho_L-\rho_R}\,
\tb^2_L\,\t^2_R\,  
{\Omega}_c \,,\\  
 {\cal U}_{v} &=v_{\a\bd}\,
\theta^\a_L \,\tb^{\bd}_R\,
e^{\rho_L-\rho_R} \,\tb_L^2\,\t^2_R \, \Omega_c \,.
\end{split}
\end{equation}
The scalar $l_{++}$ parameterizes the fluctuations
$h_{ij}=g_{j\bar\jmath}h_i^{\bar\jmath}$ of the pure components of the
internal graviton $g_{ij}$.  The complex polarization $v_{m}$ is
related to the internal components of the RR 3-form as $C_{ij\bar k} =
C \, Y_{ij\bar k}$. The complex scalar $C$ can be expressed by the
real scalar $l_{+-}$ and the dual of a real two-form field strength
$H_{mnp}$ such that $v_m=\partial_m C =(\partial_m l_{+-} +i
\epsilon_{mnpq} H^{npq})$. Reducing the RNS vertex operators for the
type IIA RR three-form and of the internal graviton one finds, in
agreement with (\ref{eMapNS}) and~(\ref{eMapR}),
\begin{equation}\label{eVOPRNStwoone}
  \begin{split}
{\cal U}_{l_{++}} &= l_{++}\, | c\,\xi e^{-\phi}|^2\;\breve\Omega_c \,,\\
 ({\cal U}_{v})^{\dagger}  &= \bar v_{\ad\b}\, \bar
S^{\ad}_L \, S_R^{\b} \,| c\,\xi\,e^{-{\phi\over2}}|^2\;\bar{\breve{\Omega}}_c\,.  
  \end{split}
\end{equation}
As for the K\"ahler moduli, the operator ${\cal U}_v$ maps
to a RNS vertex operator in a non-canonical ghost-picture.


\end{document}